\documentclass[journal]{IEEEtran}

\usepackage{xcolor,soul,framed} 
\usepackage{microtype}
\colorlet{shadecolor}{yellow}
\usepackage[pdftex]{graphicx}
\usepackage{float}

\graphicspath{{../pdf/}{../jpeg/}}
\DeclareGraphicsExtensions{.pdf,.jpeg,.png}

\usepackage[cmex10]{amsmath}
\usepackage{array}
\usepackage{mdwmath}
\usepackage{mdwtab}
\usepackage{eqparbox}
\usepackage{booktabs}
\usepackage{multirow}
\usepackage{url}
\usepackage{braket}
\usepackage{amssymb}
\usepackage[utf8x]{inputenc}
\usepackage[justification=centering]{caption}
\usepackage{ulem}				
\usepackage{quantikz}  	
\usepackage{tikz}
\usepackage{adjustbox}
\usepackage{subcaption}
\usepackage{algorithm} 
\usepackage{algpseudocode} 
\usepackage[margin=0.7in]{geometry}
\usepackage{listings}
\lstdefinestyle{lststyle}{
  commentstyle=\color{green},
  keywordstyle=\color{magenta},
  numberstyle=\tiny\color{gray},
  stringstyle=\color{purple},
  basicstyle=\ttfamily\footnotesize,
  breakatwhitespace=false,
  breaklines=true,
  captionpos=b,
  frame=lines,
  keepspaces=true,
  numbers=left,
  numbersep=5pt,
  showspaces=false,
  showstringspaces=false,
  showtabs=false,
  tabsize=2
}
\usepackage{listings, xcolor}

\definecolor{verylightgray}{rgb}{.97,.97,.97}

\lstdefinelanguage{Solidity}{
    keywords=[1]{anonymous, assembly, assert, balance, break, call, callcode, case, catch, class, constant, continue, constructor, contract, debugger, default, delegatecall, delete, do, else, emit, event, experimental, export, external, false, finally, for, function, gas, if, implements, import, in, indexed, instanceof, interface, internal, is, length, library, log0, log1, log2, log3, log4, memory, modifier, new, payable, pragma, private, protected, public, pure, push, require, return, returns, revert, selfdestruct, send, solidity, storage, struct, suicide, super, switch, then, this, throw, transfer, true, try, typeof, using, value, view, while, with, addmod, ecrecover, keccak256, mulmod, ripemd160, sha256, sha3}, 
    keywordstyle=[1]\color{blue}\bfseries,
    keywords=[2]{address, bool, byte, bytes, bytes1, bytes2, bytes3, bytes4, bytes5, bytes6, bytes7, bytes8, bytes9, bytes10, bytes11, bytes12, bytes13, bytes14, bytes15, bytes16, bytes17, bytes18, bytes19, bytes20, bytes21, bytes22, bytes23, bytes24, bytes25, bytes26, bytes27, bytes28, bytes29, bytes30, bytes31, bytes32, enum, int, int8, int16, int24, int32, int40, int48, int56, int64, int72, int80, int88, int96, int104, int112, int120, int128, int136, int144, int152, int160, int168, int176, int184, int192, int200, int208, int216, int224, int232, int240, int248, int256, mapping, string, uint, uint8, uint16, uint24, uint32, uint40, uint48, uint56, uint64, uint72, uint80, uint88, uint96, uint104, uint112, uint120, uint128, uint136, uint144, uint152, uint160, uint168, uint176, uint184, uint192, uint200, uint208, uint216, uint224, uint232, uint240, uint248, uint256, var, void, ether, finney, szabo, wei, days, hours, minutes, seconds, weeks, years}, 
    keywordstyle=[2]\color{teal}\bfseries,
    keywords=[3]{block, blockhash, coinbase, difficulty, gaslimit, number, timestamp, msg, data, gas, sender, sig, value, now, tx, gasprice, origin},   
    keywordstyle=[3]\color{violet}\bfseries,
    identifierstyle=\color{black},
    sensitive=false,
    comment=[l]{//},
    morecomment=[s]{/*}{*/},
    commentstyle=\color{gray}\ttfamily,
    stringstyle=\color{red}\ttfamily,
    morestring=[b]',
    morestring=[b]"
}

\begin{document}

\bstctlcite{IEEEexample:BSTcontrol}
    \title{Automatic Generation of Grover Quantum Oracles for Arbitrary Data Structures}
    \author{Raphael Seidel, Colin Kai-Uwe Becker, Sebastian Bock, Nikolay Tcholtchev, Ilie-Daniel Gheorge-Pop and Manfred Hauswirth\\
		Fraunhofer Institute for Open Communication Systems (FOKUS) \\ Berlin, Germany \\ \{firstname.lastname\}@fokus.fraunhofer.de}


\pagestyle{plain}
\pagenumbering{arabic}

\maketitle
\thispagestyle{plain}

\let\oldref\ref
\renewcommand{\ref}[1]{(\oldref{#1})}

\begin{abstract}
The steadily growing research interest in quantum computing - together
with the accompanying technological advances in the realization of quantum hardware - fuels the development of meaningful real-world 
applications, as well as implementations for well-known quantum algorithms.
One of the most prominent examples till today is Grover's algorithm, which can be used for efficient search in unstructured databases.
Quantum oracles that are frequently masked as black boxes play an important role in Grover's algorithm.
Hence, the automatic generation of oracles is of paramount importance. Moreover, the automatic generation of the corresponding circuits for a Grover quantum oracle is deeply linked to the 
synthesis of reversible quantum logic, which - despite numerous advances in the field - still remains a challenge till today in terms of 
synthesizing efficient and scalable circuits for complex boolean functions. 

In this paper, we present a flexible method for automatically encoding unstructured databases into oracles, 
which can then be efficiently searched with Grover's algorithm. Furthermore, we develop a tailor-made method for 
quantum logic synthesis, which vastly improves circuit complexity over other current approaches. Finally, we present another logic synthesis method that considers the requirements of scaling onto real world backends.
We compare our method with other approaches through evaluating the oracle generation for random databases and analyzing the resulting circuit 
complexities using various metrics.
\end{abstract}


%
\IEEEpeerreviewmaketitle

\label{sec:introduction}

\section{Introduction}

\subsection{General Remarks}
The field of quantum computing has seen a significant rise in interest over the past two years since the quantum supremacy announcement from Google \cite{Arute} and will continue to be of great interest, among other things, due to the recent supremacy announcements regarding chinese quantum hardware designs \cite{Zhong2020}, \cite{Wu2021}, \cite{Zhu2021}. Yet some of the well known algorithms that are representative for the potential of quantum computing are still far from being implementable for meaningful applications on today’s quantum hardware. Especially noteworthy are Shor's algorithm that could play a significant role in cryptography and cybersecurity as well as Grover's algorithm. The latter was initially designed for 
searching through large unstructured databases \cite{Grover1996} but also has applications in cryptography \cite{Sun2014} (e.g. for the 
search of encryption keys in the context of symmetric cryptographic algorithms), optimization problems \cite{Gilliam_2021, Chakrabarty2017, 
Baritompa2005} - that can even be beneficial for material discovery \cite{Borujeni2019} - as well as applications for challenging problems 
in the domain of data structure and hash function design \cite{Brassard1998}. 

The focus of the research presented in this paper relates to the scalable synthesis of quantum oracles used in the context of Grover 
database search. The need for automatic oracle generation for Grover's algorithm was extensively discussed in one of our previous 
research works \cite{GheorghePop2021}. After a thorough analysis, \cite{GheorghePop2021} clearly states that there is a paramount requirement for automatically generated oracles as inputs for Grover's algorithm, in order to be able to provide
function/procedure APIs (Application Programming Interfaces) to developers and that way seamlessly integrate quantum computing
into existing software development projects (e.g. telecom supporting systems, web shops and other kinds of applications) and 
processes. In order to address this urgent need, we present the design and implementation of a fully automated programming 
framework for the complete Grover based search process, including the generation of the belonging quantum oracles. This involves the conversion of arbitrary data into boolean functions that are encoded in truth tables, 
the synthesis of Grover oracles from these truth table expressions and the direct integration into the initialization and 
diffusion steps of Grover's algorithm.

\subsection{Problem Statement}
The following constitutes a tangible list of aspects that summarize the targeted problem:
\begin{itemize}
    \item The creation of oracles for Grover's search algorithm is a challenging task.
    \item The creation of quantum oracles containing the input for a Grover search needs to be done on a case-by-case basis
		depending on the database/list to search in and the object/value to search for.
		\item As of the current state-of-the art there is no accessible computer scientist's alike API for functions/procedures 
		(e.g. \textit{int grover(int [] list\_to\_search\_in, int value\_to\_search\_for)})
		that allows for effortless submission of a database/list and an object/value to search for in this database/list.
		\item Hence, Grover's algorithm is almost impossible to apply for the execution of standard tasks in the course
		of programing IT components and products in different application domains
\end{itemize}

In order to address the above issues, our current research paper presents a solution, such that efficient and convenient I/O procedures to Grover's algorithm are enabled. The following
sub-section outlines the specific contributions of our paper which address the posed problem statement and belonging research challenge.

\subsection{Contributions}
The following key contributions are the result of our research work, which is presented in the current paper
towards progressing on the topic of automatic generation of Grover quantum oracles for arbitrary data structures:

\begin{enumerate}
    \item The current paper presents an overall structure of a procedure for the automatic generation of Grover oracles, which
		is based on truth tables and belonging quantum logic synthesis. This procedure enables Grover quantum search of arbitrary databases.
    \item We present a high level complexity analysis of this general structure/procedure and clearly identify the benefits and 
		the need for improvements of the algorithms and the approach itself.
		\item Subsequently, we present a flexible method for implementing similarity searches over a database thereby steering the oracle
		generation correspondingly. Within a similarity search it is possible to utilize Grover's algorithm to identify database items
		which have a certain degree of resemblance compared to a particular item we are searching for.
		\item Moreover, we survey the topic of quantum logic synthesis and adapt different available algorithms - such as the 
		\textit{Reed-Muller Expansion} and the \textit{Gray Synthesis} in order to improve their efficiency in the workflow of oracle generation.
		\item We present a new \textit{Phase Tolerant Synthesis} method, a streamlined version of the established Gray Synthesis, which (asymptotically) halves the required resources.
		\item Another critical issue is constituted by the scaling of logic synthesis methods which quickly requires extremely fine phase gates. In order to address this issue we introduce the method of CSE-synthesis.
		\item Finally, we present a number of benchmarking results which confirm the efficiency of our approaches and algorithms for 
		quantum logical synthesis and in general for automatic Grover oracle generation.
\end{enumerate}

The above items constitute clear contributions, which to the author's knowledge are the first attempt so far to enable the efficient
and convenient input of databases and objects to search for with Grover's algorithm with the goal to make it usable for different tasks in modern IT components and products (e.g. web portals, search engines, network operation centers etc.).

\subsection{Structure of the Paper}
The rest of this paper is organized as follows: In section \ref{overview} we present an overview of the state-of-the-art in oracle 
synthesis work. In the subsequent section \ref{method} our method for automatic oracles generation is presented. In section \ref{synthesis}, we give an introduction into different methods for reversible quantum logic synthesis and 
elaborate on our improved version for Grover oracle generation that we denote as \textit{phase tolerant synthesis}. This is then 
further discussed in the context of scalable logic synthesis circuit design in section \ref{performance_optimization}. 
Section \ref{benchmarking} describes the tests and results we obtained by comparing existing oracle generation methods with the one we 
developed in terms of circuit complexity. Finally, section \ref{summary} contains a summary of the research presented in this paper as well 
as some brief discussion about our future research plans.

\label{sec:overview}
\section{Overview and Background}
\label{overview}

To provide an initial overview, section \ref{overview} reviews Grover's algorithm for efficient item search in unstructured databases on a higher level without
going into the mathematical details and proofs. Afterwards, some recent research works from the domain of
\textit{Reversible Quantum Logic Synthesis} are presented. These activities and related results are of particular importance
for our approach to the automated generation of Grover quantum oracles.
Besides specific applications for Grover's algorithm, proof of concepts for experimental realizations \cite{Samsonov2020} and low-scale 
prototype implementations on current quantum hardware \cite{Mandviwalla2018}, a really important field of research lies in the scalable 
generation of Grover oracles and quantum oracles for black box algorithms in general \cite{GheorghePop2021}.

\subsection {Grover's Algorithm}
 At the heart of Grover's algorithm lies a quantum oracle, which phase-tags only the winner items. For now we leave the specifics of generating said oracle from the database to the subsequent sections and simply assume it has the following functionality: 
\begin{align}
\label{oracle_equation}
O\ket{i} = (-1)^{f(i)} \ket{i} \ \ \text{with} \ \ f(i) = \begin{cases} 0, & \text{if } i \not\in \{w_j\}\\
1, & \text{if } i \in \{w_j\}.
\end{cases}
\end{align}
Note that for the oracle, we don't need to explicitly know $w_j$ but we only need a valid function $f$ for the search problem. Also for an 
explicit construction of such oracles, one generally needs an auxiliary qubit register to store intermediate results into which has to be uncomputed later on\footnote{\textit{Uncompation} is the process of \textit{Garbage Collection} with 
respect to qubits, i.e. during an uncomputation the corresponding qubits are systematically reset to their initial states through the 
belonging reversible mathematical operations.}. The role of this auxiliary qubit register and belonging operations is further
discussed in section \ref{method}.

In order to evaluate Grover's algorithm with this oracle, we initialize the system in the fiducial state $ \ket{\Psi} = \ket{0}^n$. Grover's algorithm starts 
by setting all qubits into an equal superposition state $\ket{s}$
\begin{align}
H^{\otimes n}\ket{0}^n = \frac{1}{\sqrt{N}} \sum_{i=0}^{N-1} \ket{i} = \ket{s}.
\end{align}
Here, the integer states $\ket{i}$ directly relate to corresponding binary encoded states in the computational basis.

After phase-tagging the winner states - i.e. the states representing the values we search for in the unstructured database - Grover's 
algorithm implements a so called diffusion operator $U_d = 2\ket{s}\bra{s} - \mathbb{I}$ that amplifies the amplitudes for measuring the 
winner states. The phase-tagging and diffusion steps can geometrically be considered as two successively performed reflections, and thus as a 
single rotation in a 2D-plane. Each such rotation corresponds to a single Grover iteration that 
gradually rotates $\ket{s}$ closer to $\ket{w_j}$. At the end of the algorithm, a measurement in the computational basis is performed and 
the searched and potentially found items can be identified by distinct peaks in the distribution of the measured results. 
It is important to remark that in literature \cite{Bennett_1997} there is a derived optimal number of Grover 
iterations (i.e. amplification and oracle application) that results in $\ket{s}$ being rotated the closest to $\ket{w_j}$. This optimal 
number of iterations yields the best measurement results. Thus, performing more 
or less rotations is expected to lead to increasingly worse search results. A more in-depth discussion about Grover iterations will 
be given in section \ref{method}. To summarize briefly: searching $M$ items within an unstructured database needs 
at most $\mathcal{O}(\sqrt{\frac{N}{M}})$ iterations by Grover's algorithm. Hence there is a noticeable advantage over classical search algorithms, which are known to perform in $\mathcal{O}(N)$ for a linear search.

\subsection {Reversible Quantum Logic Synthesis}
 Technically, there is a deep 
interconnection between generating quantum oracles and synthesizing reversible quantum circuits from classical logic expressions. 
Therefore, in section \ref{method} we describe the application of state-of-the-art quantum logic synthesis procedures for our approach to 
Grover oracle synthesis. In the following paragraph, we conduct a brief introduction to related research and development activities from
the area.

As \textit{Reversible Quantum Logic Synthesis} is an active field of research for more than 20 years \cite{Anas_2004}, 
more efficient methods in terms of gate counts and the usage of ancilla qubits arise frequently \cite{PhysRevA.85.044302}. 
While many of them are available in the \textit{tweedledum} \cite{tweedledum} library, a fairly recent method 
of interest is denoted as \textit{resource-efficient oracle synthesis (ROS)} and was initially presented in \cite{Meuli2020}. It is built 
on the \textit{LUT\footnote{LUT stands for \textit{lookup-table}.}-based hierarchical reversible logic synthesis framework} 
(LHRS) \cite{Soeken2017} and improves the LUT-network approach 
by adding a special quantum aware k-LUT mapper, involving specially truncated XAGs\footnote{XAG stands for Xor-And-inverter 
Graphs, which constitute a particular form of representing intermediate results during the process of quantum logic synthesis.} and a more 
efficient SAT-based quantum garbage management technique\footnote{A management technique for the efficient uncomputation (i.e. 
garbage collection) of the corresponding circuits involving instances of the SAT-problem (SAT=SATISFIABILITY)}  \cite{Meuli2019}, which 
offers control over the number of qubits used in the 
synthesized ciruits. This method can be considered as a reasonable competitor and thus constitutes an 
interesting research work for comparison to our synthesis methods. Additionally, there are some implementations for oracle generation in the 
established software frameworks like Qiskit \cite{qiskit_TruthTableOracle} and Q\# \cite{q_sharp_OracleSynthesis}. 
The Qiskit implementation will be tested and compared to our approach in section \ref{sec:benchmarking}.
Apart from the above mentioned activities, there are no further comparable methods to the best of our knowledge.


\label{sec:method}
\section{Method for Automatic Oracle Generation}
\label{method}
We will now present our method of turning arbitrary databases into oracles. 
The basic idea is to deploy a computationally effective labeling function, which turns the data into bitstrings of suited length. 
For a database $D$, this labeling will be denoted as\footnote{$\mathbb{F}_2$ is the Galois field with two elements, i.e. the field of operation 
for traditional boolean algebra. The arithmetic operations are all carried out $\bmod 2$, i.e. $1+1 \equiv 0 \bmod 2$. We will use the $
\oplus$ notation for additions in the $\mathbb{F}_2$ space.}

\begin{align}
	l: D \rightarrow \mathbb{F}_2^k \\ 
	e \rightarrow l(e) \nonumber.
\end{align}

Here, $k \in \mathbb{N}$ stands for the label size, which has to be chosen according to the specifics of the problem (more to that in section \ref{label_hash_collisions}). 
The method of obtaining these bitstrings is in principle also free to choose as long as it only takes local information from the database, 
i.e. the bitstring of each element is independent from the rest of the database. 
For example, in our Python implementation, the native \textit{hash} function generates an integer hash value for a wide class of objects, which 
we convert to the belonging binary representation and clip the bitstring to a length of $k$.

As a result, the labels provide a sequence of bitstrings representing the database, 
which eventually can be used to form a logical truth table (see table \ref{ex_tt} for an example). 
Subsequently, this truth table is turned into a quantum circuit using a suited quantum logic synthesis algorithm. 
Given two registers (the index and the label register) this yields a unitary mapping $U_D$ which acts as

\begin{align}
\label{database_circuit}
U_D\ket{i}\ket{0} = \ket{i}\ket{l(e(i))}.
\end{align}
where $l(e(i))$ is the label of the database entry $e$ with index $i$. It is important to note that the synthesis process only has to be 
done once per database. The resulting circuit can then be used for subsequent searches and only needs to be updated at changes to the database entries. We leave the challenge of efficient updating for database circuits as an open research question, which we are going to address in the near future.

Based  on the above considerations, in order to \textbf{q}uery the index $i_q$ for a given database element $e_q$, we combine the synthesized truth table with a phase-tag of the bitstring $l(e_q)$, which can be calculated with very small effort from $e_q$. The phase-
tag is realized using a multi-controlled $Z$ gate which is enclosed by $X$ gates at the appropriate qubits. Finally, the label variable 
has to be uncomputed again. Denoting the tagging function of label $l(e)$ with $T(l(e))$, the mathematical description of the actions of 
the oracle $ O(e_q) = U_D^\dagger T(l(e)) U_D$ of the element $e_q$ is as follows:

\begin{equation}
	\begin{aligned}
	&O(e_q) \ket{i}\ket{0}\\
	=& U_D^\dagger T(l(e_q)) U_D \ket{i}\ket{0} \\
	=& U_D^\dagger T(l(e_q)) \ket{i}\ket{l(e(i))}\\
	=& \begin{cases} -U_D^\dagger \ket{i}\ket{l(e(i))} & \text{if $l(e_q) = l(e(i))$}\\
	U_D^\dagger \ket{i}\ket{l(e(i))} & \text{else} \end{cases}\\
	=& \begin{cases} -\ket{i}\ket{0} & \text{if $l(e_q) = l(e(i))$}\\
	\ket{i}\ket{0} & \text{else} \end{cases}
	\end{aligned}
\end{equation}
which is precisely the functionality we required $O(e_q)$ to have in eq. \ref{oracle_equation}.
\subsection{Hash collisions}
\label{label_hash_collisions}
\begin{figure*}
\begin{subfigure}{.45\textwidth}
\begin{tabular}{c|ccc|c|cccc}
$i$ & $i_0$ & $i_1$ & $i_2$ & $e$ & $l_0(e)$ & $l_1(e)$ & $l_2(e)$ & $l_3(e)$\\
\hline
0 & 0 & 0 & 0 & Alice& 1 & 1 & 0 &0\\
1 & 0 & 0 & 1 & Bob& 0 & 1 & 0 &1\\
2 & 0 & 1 & 0 & Craig& 0 & 0 & 1 &1\\
3 & 0 & 1 & 1 & Dan& 1 & 1 & 0 &1\\
4 & 1 & 0 & 0 & Eve& 0 & 0 & 0 &1\\
5 & 1 & 0 & 1 & Faythe& 0 & 0 & 1 &0\\
6 & 1 & 1 & 0 & Grace &0 & 1 & 0 &1\\
7 & 1 & 1 & 1 & Heidi& 1 & 0 & 0 &1\\
\end{tabular}
\caption{\label{ex_tt}}
\end{subfigure}
\begin{subfigure}{.45\textwidth}
\includegraphics[width = \textwidth]{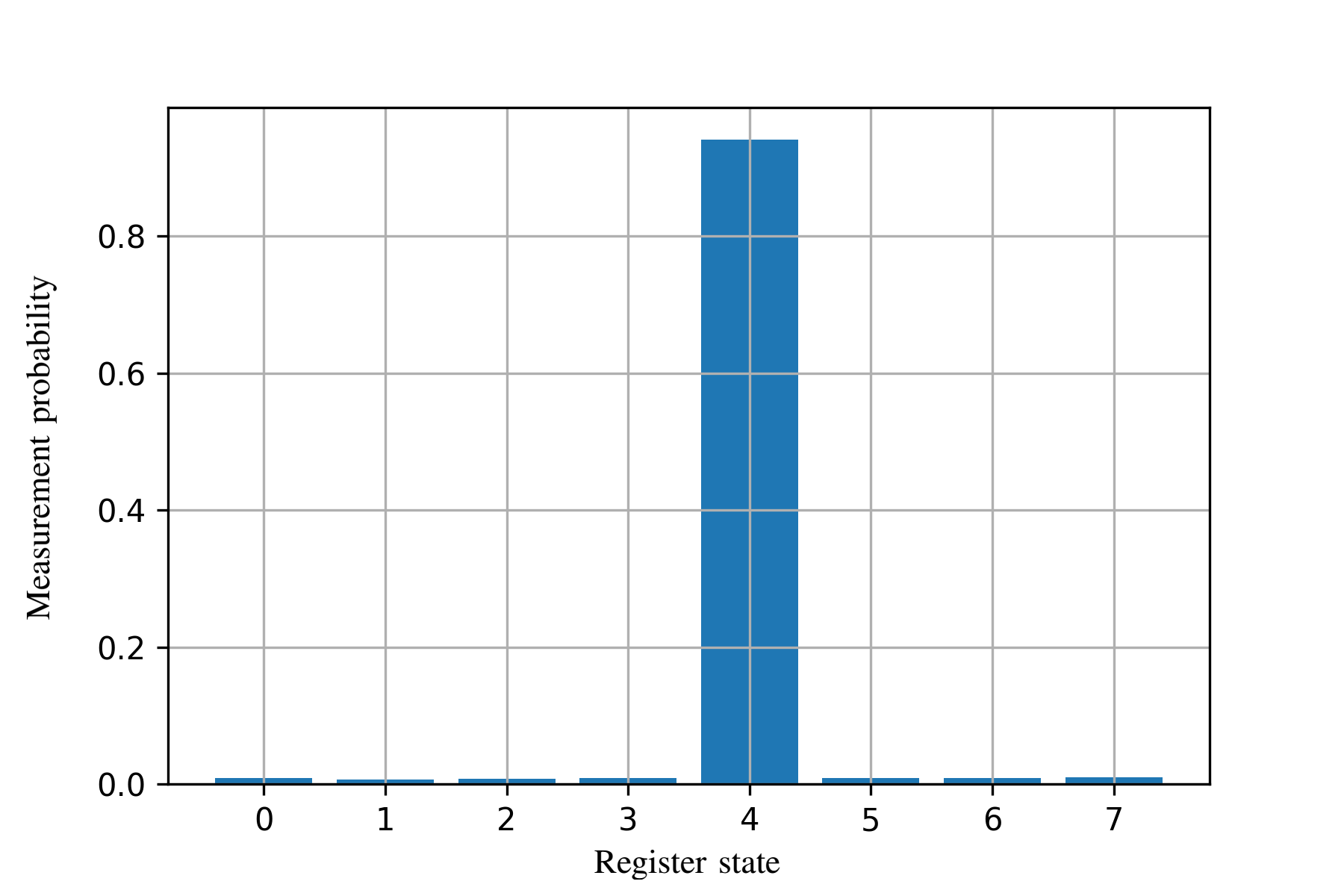}
\caption{\label{grover_example}}
\end{subfigure}
\begin{subfigure}{.5425\textwidth}
\begin{quantikz}
\lstick[wires = 3]{$(H\ket{0})^{\otimes n}$}  &\gate[6, disable auto height]{U_D}&\qw & \gate[6, disable auto height]{U_D^\dagger} & \gate[3]{\text{Diffuser}} & \qw \\
&&\qw && &\qw \\
&&\qw && &\qw \\
\lstick[wires = 3]{$\ket{0}^{\otimes k}$} && \gate[wires = 3]{T_{l(e_q)}} &&\qw &\qw &  \\
&&&&\qw &\qw & \\
&&&&\qw & \qw & 
\end{quantikz}
\caption{\label{grover_iteration}}
\end{subfigure}
\begin{subfigure}{.45\textwidth}
\includegraphics[width = \textwidth]{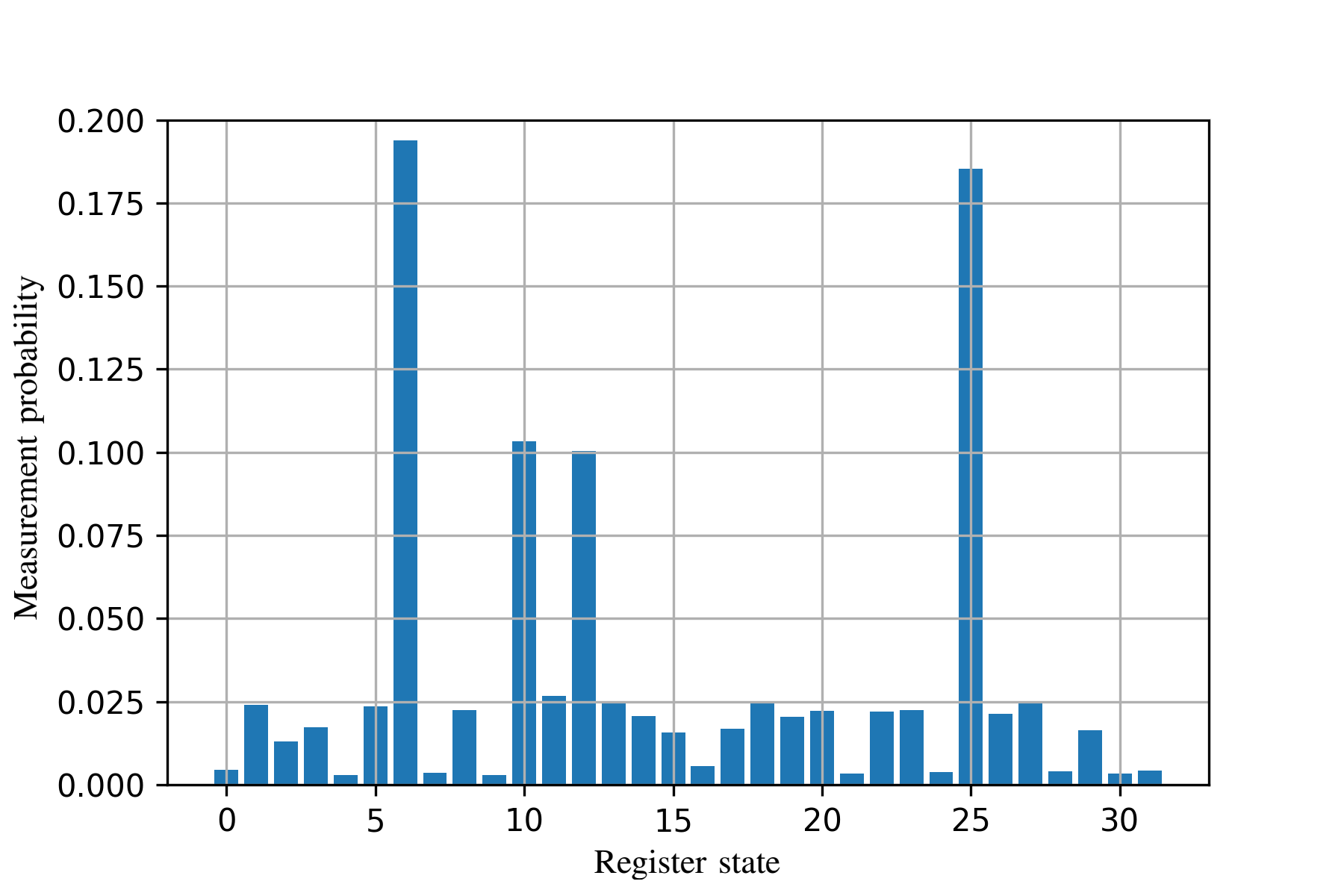}
\caption{\label{grover_sim_example}}
\end{subfigure}
\caption{ \textbf{\ref{ex_tt}} 
Truth table of an example database containing some names as entries. The values for $l(e)$ are generated by truncating the binary strings 
from the output of Python's native \textit{hash} function. \textbf{\ref{grover_example}} Example application of Grover's algorithm to the oracle created from the database in table 
\ref{ex_tt} querying for "Eve". The simulation was performed with \textit{Qiskit}s OpenQASM simulator \cite{Qiskit}. \textbf{\ref{grover_iteration}} Quantum circuit of a single grover iteration quering for the database element $e_q \in D$.\textbf{\ref{grover_sim_example}} Histogram of the measurement probabilities 
after applying Grover's algorithm to an oracle which uses the tagging gate from eq. \ref{similarity_tag}. As query label we used the 
label of the element at index 6 which is 110011. The label from index 10 is 010011 and the label at index 12 is 111011 - both of these 
have Hamming distance 1 from the query label. }
\end{figure*}

As can be seen in table \ref{ex_tt}, it can happen that two elements have the same label (here: Grace \& Bob). This poses not that much of 
a problem, since after running Grover's algorithm a classical search can be applied on the (heavily reduced) result space. Another insight is 
constituted by the fact that the probability for a hash collision is well controlled by increasing/decreasing the label 
size (i.e. the size of the binary string). A second maybe more subtle problem is that in order to determine the optimal amount of Grover iterations $R$, the amount of tagged states has to be known according to \cite{Chen1999}

\begin{align}
R \leq \left\lceil \frac{\pi}{4} \sqrt{\frac{N}{M}} \right\rceil
\label{eqn:grover_iterations}
\end{align}

with $M$ being the amount of elements sharing the labeling bitstring $l(e_q)$ and $N$ the amount of elements contained in the database $D$. This might 
seem only like a minor inconvenience, because on first sight this could result in only a few extra-iterations. However, as described in 
the previous section, the results get worse if continuing with iterations after the optimal prescribed number. 
A possible quantum algorithm, which can determine $M$, is presented in \cite{Brassard1998_counting}. Unfortunately, it requires 
exponentially many oracle calls, which is untenable for an efficient database and fast system reaction times. In addition, 
all these oracle calls have to be controlled, which implies that every CNOT gate is turned into a Toffoli gate \footnote{A Toffoli gate is a a controlled CNOT gate (or CCNOT). This gate can be synthesized using 6 CNOT gates.} leading to another factor of 6 in the CNOT count. Due to these disadvantages, the approach is not feasible in practice, which is why we use a heuristic approach to 
estimate $M$.
In detail, we determine the expected value of another element $e_i$ colliding with the particular bitstring of $e_0$. For this we assume 
that the bitstrings are uniformly distributed over the label space $\mathbb{F}_2^k$. This is the setup for a Bernoulli-Process with $n = N-1$ 
tries with a probability of $p = 2^{-k}$. The resulting model is constituted by the binomial distribution and the expected value 
therefore takes the very simple form of
\begin{align}
\mathbb{E}(\text{\#collisions}) = np = (N-1)2^{-k}.
\end{align}
Since there is at least one element sharing the bitstring of $e_0$ (i.e. $e_0$ itself) we add a $1$ to acquire the expected value for $M$:
\begin{align}
\mathbb{E}(M) = 1 + (N-1)2^{-k}.
\end{align}

This formula captures the intuitive relationship between the label size $k$ and the number
of estimated hash collision $M$. It shows that by decreasing the label size $k$, one can increase $M$, while in  
parallel decreasing the amount of Grover iterations (see eq. \ref{eqn:grover_iterations}) 
to be executed and synthesis tasks to be performed. This would be interesting in future hybrid use cases, where quantum resources are 
available at a similar cost as classical resources but still not capable of solving complex problems on their own.
In this hybrid scenario, the search space would first be drastically lowered by a factor of $\mathcal{O}(2^{-k})$ by the quantum computer 
and then reduced to $1$ with a classical search. 

\subsection{Algorithmic View on Oracle Generation}
\label{Algorithmic_View_on_Oracle_Generation}
We want to continue by summarizing the overall procedure for the automated oracle generation from an algorithmic point of view. Thereby, we provide an overall structure of the steps which are required to generate an oracle 
for Grover's algorithm from a random database. These are structured in two sub-algorithms,
namely the one responsible for database encoding (i.e. algorithm \ref{alg:database_encoder}) and 
the second one, which is responsible for automatic definition of the oracle based on the encoded 
database (i.e. algorithm \ref{alg:query_oracle_genarator}). Based on those two sub-algorithms, time 
complexity estimations are provided, in order to give a feeling regarding the applicability and the open research issues relating to the topic of automatic oracle generation.

\begin{algorithm}
	\caption{Database encoder} 
	\label{alg:database_encoder}
	\hspace*{\algorithmicindent}\textbf{Input:} Iterable database $D$, labeling function\\
	\hspace*{\algorithmicindent}\textbf{Output:} Encoding circuit $U_D$
	\begin{algorithmic}[1]
		\State list bitstrings = []
		\For {$e$ in $D$}
			\State bitstrings.append(Label(e))
			\EndFor
		\State TruthTable tt = TruthTable(bitstrings)
		\State \textbf{return} QuantumLogicSynthesis(tt)
	\end{algorithmic} 
\end{algorithm}

Algorithm \ref{alg:database_encoder} describes the steps required to encode an arbitrary
database for the purposes of Grover's search algorithm. The only requirement here is, that it should be possible to iterate over the content of the database.
Thereby, no order of the content is presumed. Hence, the database
can contain arbitrary objects. 

The other ingredient, which is required as an input to algorithm \ref{alg:database_encoder},
is the \textit{labeling function}. This function needs to have the capability 
to take an arbitrary object from the database and assign it a label, that does not have to be unique.In fact, hash functions as known in computer science are natural choices as labeling functions. Hash functions in general do not provide unique labels (i.e. hash collisions). However, even with non-unique labels of the
database objects, Grover search can be used to search for labels and 
that way reduce the overall search complexity by collapsing the search space only
to those objects/entries, which have been assigned a particular label.

As can be seen in the listing of algorithm \ref{alg:database_encoder}, a \textit{for}-cycle
is required that iterates over the database objects and assigns a label to each of them.
This cycle has the time complexity of $\mathcal{O}(N)$, with $N$ being the number of entries in the database.
Each label is presumed to be a binary string and is stored in a corresponding list, which is used to initially prepare a truth table for Grover oracle. This truth table
consists of simply listing and preparing the label bitstrings in such a way they can be used for quantum logic synthesis. Hence, the truth table creation has again complexity of $\mathcal{O}(N)$ and and leaves the previous complexity estimation unchanged.

The quantum logic synthesis in the last step of algorithm \ref{alg:database_encoder} is the most 
critical part of the process. This step can be conducted by various algorithms and procedures, which is 
further focused on in the following sections of this paper, in which some fitting methods from literature
are applied and evaluated and subsequently own methods are presented. As we will see in the upcoming section \ref{synthesis} this step can be performed quite efficiently using an algorithm called \textit{Fast Hadamard-Walsh Transform} \cite{Fino1976} which has classical time complexity $\mathcal{O}(N \times \text{log}(N))$ per truth table column. In our case we synthesize about $k \approx \mathcal{O}(\text{log}_2(N))$ in order to prevent excessive hash collisions, yielding an classical time complexity of $\mathcal{O}(N \times \text{log}(N)^2)$.

However, as algorithm \ref{alg:database_encoder} has to be executed only once per fixed database, Grover's algorithm can be executed for different values to search for on the prepared database quantum circuit. Thus, the time complexity can be relieved this way with regard to the application of algorithm \ref{alg:database_encoder}.

\begin{algorithm}
	\caption{Query oracle generator} 
	\label{alg:query_oracle_genarator}
	\hspace*{\algorithmicindent}\textbf{Input:} Encoded database circuit $U_D$, query element $e_q$, labeling function\\
	\hspace*{\algorithmicindent}\textbf{Output:} Oracle circuit $O$
	\begin{algorithmic}[1]
		\State string qlabel = Label($e_q$)
		\State QuantumCircuit $O$ = QuantumCircuit(index register, label register)
		\State $O$.apply($U_D$)
		\State $O$.apply(PhaseTag(qlabel, label register))
		\State $O$.apply($U_D^{\dagger}$)
		
		\State \textbf{return} $O$
		
	\end{algorithmic} 
\end{algorithm}

After the steps from algorithm \ref{alg:database_encoder} have been conducted, algorithm \ref{alg:query_oracle_genarator}
must be executed every time before a new object/value is being searched for in the database entries. 
Algorithm \ref{alg:query_oracle_genarator} contains the required instructions towards preparing a specific quantum oracle 
encoding the database and the search value.
This specific quantum oracle is then embedded in Grover's algorithm for the subsequent search on the quantum 
computer side. We see the following components in algorithm \ref{alg:query_oracle_genarator}: The creation of a 
$QuantumCircuit$ object, in which the quantum circuit encoding the database (result from algorithm \ref{alg:database_encoder}) is immediately embedded (i.e. O.apply($U_D$) and finally O.apply($U_D^{\dagger}$) followed by the the application of a phase tag
(i.e. O.apply(PhaseTag(qlabel,labelregister)) ) for marking the value/object to search for.
The steps in the first component are dominated by the embedding of the quantum encoded database circuit (i.e. O.apply($U_D$) and finally O.apply($U_D^{\dagger}$) which can be considered linear with respect to the number of gates in the database quantum circuit. As we will see in the upcoming sections, the Gray synthesis method has a quantum gate complexity of $\mathcal{O}(N)$ per truth table column, implying a quantum gate complexity of $\mathcal{O}(N\times \text{log}(N))$ for $k \approx \mathcal{O}(\text{log}(N))$ columns. The phase tag (i.e. O.apply(PhaseTag(qlabel, labelregister))) and the diffuser can also be synthesized using Gray synthesis i.e. $\mathcal{O}(N)$. As we require $\mathcal{O}(\sqrt{N})$ Grover iterations, the overall quantum gate complexity is 
\begin{align}
\mathcal{O}(N^{1.5}\times \text{log}(N)).
\end{align}
On first sight this seems like a decrease in Grover's efficiency $\mathcal{O}(\sqrt{N})$ however we have to keep in mind that this is the complexity in \textit{oracle calls} the oracle evaluation itself will have a non-constant complexity in almost any other comparable scenario.\\ Even though currently not more efficient than classical database search, our results open many research directions which could ultimately lead to an increase in database search efficiency.\\
Having briefly discussed this, the following sections continue with the presentation and deepen the discussion on various methods for quantum logic synthesis and for establishing Grover oracles and the search procedures in a flexible way.

\subsection{Similarity Search}

The method presented so far can be generalized to a technique, which also allows for searching bitstrings \textit{similar} to the query, 
i.e. not the exact item but rather one that is very close according to some metric. In the case, where the oracle has been generated from 
labeled data, this can obviously only work if the labeling function preserves the similarities between the database elements. Another 
application scenario for a \textit{similarity search} could be the case where the labels themselves constitute the data.

The general idea of encoding the similarity of two bitstrings into the quantum oracle is to replace the tagging function $T^{\text{sim}}(l(e_q))$ with a circuit 
that performs phase shifts based on the Hamming-Distance. One such circuit is given by the application of RZ gates on the label register. 
In more detail:
\begin{align}
\label{similarity_tag}
T^{\text{sim}}(l(e_q)) = \bigotimes_{j = 0}^{k-1} \text{RZ}_i\left(\frac{-(-1)^{l(e_q)_j} 2\pi}{k} \right).
\end{align}
For a single RZ gate acting on a qubit in a computational basis state we have
\begin{align}
RZ(-(-1)^{x} \phi) \ket{y} = \text{exp}\left(\frac{i \phi}{2}(-1)^{x\oplus y}\right) \ket{y}.
\end{align}
Applying the similarity tag $T^{\text{sim}}(l(e_q))$ to the multi-qubit state $\ket{l(e)}$ therefore yields:
\begin{align}
\label{similarity_tag_phase}
T^{\text{sim}}(l(e_q)) \ket{l(e)} = \text{exp}\left(\frac{i\pi}{k}\sum_{j = 0}^{k-1} (-1)^{l_j(e_q) \oplus l_j(e)}\right) \ket{l(e)}.
\end{align}
To get an intuition of the effect of this, we now consider what happens when $l(e) = l(e_q)$. In this case we have:
\begin{align}
 l_i(e_q) \oplus l_i(e) = 0 \ \ \forall i < k
\end{align} implying that the sum over $j$ evaluates to $k$. Therefore the applied phase is simply $\pi$, i.e. exactly what we have in the 
case of a regular phase-tag. If $l(e) = l(e_q)$ for all but one $j$-index, the sum evaluates to $k-2$. Hence, the applied phase is 
$\pi(1 - \frac{2}{k})$ which is "almost" the full phase-tag.
An example application of the similarity search can be found in fig. \ref{grover_sim_example}.

In the following, a more formal proof is presented regarding why the above considerations work when applied within Grover's algorithm.
For this we assume that the oracle has tagged the uniform superposition $\ket{s} = \frac{1}{\sqrt{N}} \sum_{x = 0}^{2^n-1} \ket{x}$ in 
such a way that the index register is in the state
\begin{align}
\label{similarity_phase_tag_eq}
\ket{\psi} = \frac{1}{\sqrt{N}} \sum_{x = 0}^{2^n-1} \text{exp}(i\phi (x)) \ket{x}.
\end{align}
We now apply the diffusion operator $U_s = 2 \ket{s}\bra{s} - \mathbb{I}$ on the above state:
\begin{equation}
\begin{aligned}
U_s \ket{\psi} &= \frac{1}{\sqrt{N}} (2 \ket{s}\bra{s} - \mathbb{I}) \sum_{x = 0}^{2^n-1} \text{exp}(i\phi (x)) \ket{x}\\
& = \frac{1}{\sqrt{N}} \sum_{x = 0}^{2^n-1} \text{exp}(i\phi (x)) (2 \ket{s}\braket{s|x} - \ket{x}).
\end{aligned}
\end{equation}
Using $\braket{s | x} = \frac{1}{\sqrt{N}}$ gives
\begin{align}
& = \frac{1}{\sqrt{N}} \left( \frac{2}{\sqrt{N}} \left( \sum_{x = 0}^{2^n-1} \text{exp}(i\phi (x)) \right) \ket{s} - \sum_{x = 0}^{2^n-1} \text{exp}(i\phi (x)) \ket{x} \right).
\end{align}
Next, we set 
\begin{align}
\label{cm_eq}
r_{\text{cm}} \text{exp}(i\phi_{\text{cm}}) := \frac{1}{N} \sum_{x = 0}^{2^n-1} \text{exp}(i\phi (x)),
\end{align}
where cm stands for the center of mass. Inserting the definition of $\ket{s}$ we get
\begin{equation}
\begin{aligned}
&= \frac{1}{\sqrt{N}} \sum_{x = 0}^{2^n-1} (2r_{\text{cm}}\text{exp}(i\phi_{\text{cm}}) - \text{exp}(i\phi(x))) \ket{x}\\
&= \frac{\text{exp}(i\phi_{\text{cm}})}{\sqrt{N}} \sum_{x = 0}^{2^n-1} (2r_{\text{cm}} + \text{exp}(i(\phi(x)-\phi_{\text{cm}} + \pi))) \ket{x}.
\end{aligned}
\end{equation}
Finally, we use the rules of polar coordinate addition
\begin{align}
r_3 \text{exp}(i\phi_3) = r_1 \text{exp}(i\phi_1) + r_2 \text{exp}(i\phi_2),\\
r_3 = \sqrt{r_1^2 + r_2^2 + 2r_1 r_2 \text{cos}(\phi_1 - \phi_2)}\\
\phi_3 = \text{arctan}\left(\frac{r_1 \text{sin}(\phi_1) + r_2 \text{sin}(\phi_2)}{r_1 \text{cos}(\phi_1) + r_2 \text{cos}(\phi_2)} \right),
\end{align}
to determine the absolute values of the coefficients in order to find out about the amplification factor $A_x$
\begin{equation}
\label{amplification_eq}
\begin{aligned}
A_x &= |2r_{\text{cm}} + \text{exp}(i(\phi(x)-\phi_{\text{cm}} + \pi))|\\
&= \sqrt{1 + 4r_{\text{cm}}^2 - 4r_{\text{cm}} \text{cos}(\phi(x)-\phi_{\text{cm}})}.
\end{aligned}
\end{equation}

From this we see that $A_x$ becomes maximal if $\phi(x) - \phi_{\text{cm}} = \pm \pi$, minimal if $\phi(x) - \phi_{\text{cm}} = 0$ and is monotonically 
developing in between, which is precisely the behavior we expected.\\
Even though the similarity tag allows a considerable cut in the CNOT count 
compared to the multi-controlled $Z$ gate, it comes with some drawbacks.
The biggest problem is that the results of this method are very sensitive to the number of iterations. Applying the wrong amount of Grover 
iterations can lead to the case, where labels which are less similar get a higher measurement probability. This might be improved through 
further further analysis of the similarity phase-tags.
Another drawback is that labels, where the bitwise NOT is similar to the query, get the same amplification as their inverted counter-part 
(assuming $\phi_{\text{cm}} = 0$ in eq. \ref{amplification_eq} - more to this assumption soon). 
For example if we query for the label 000000, the label 011111 would receive the same amplification as 100000. This can be explained
by looking at equation \ref{similarity_tag_phase}. If $l(e_q)$ and $l(e)$ differ on every single entry, we have
\begin{align}
 l(e_q)_i \oplus l(e)_i = 1 \ \ \forall i < k
\end{align}
This results in the sum evaluating to $-k$ and yielding a phase of $-\pi$, which is equivalent to a phase of $\pi$.
A similar effect can be observed, when only a single bit of the inverse is mismatching. In this case, we apply the 
phase $-\pi +\frac{1}{k}$. Even though this phase is not equivalent to $\pi -\frac{1}{k}$, its absolute value is, which is the  
feature of relevance according to equation \ref{amplification_eq}.\\

\subsection{Advanced Similarity Tags}
In this section we will present a method which lifts the restriction of the similarity tags being confined to the Hamming 
distance. Instead, it will be possible to encode an extremely wide and flexible class of similarity measures (without any 
overhead compared to the regular multi-controlled Z gate tags). If we denote the set of possible query objects with $Q$, any 
function with the following signature can be encoded as a similarity measure:

\begin{align}
\label{similarity_measure_eq}
f: Q \times \mathbb{F}_2^k \rightarrow [0,1].
\end{align}
Here $\mathbb{F}_2^k$ again denotes the set of bitstrings with length $k$ and $[0,1]$ the interval between 0 and 1 (endpoints 
included). So this function compares a query object $q \in Q$ with a bitstring and returns a real number between 0 and 1, 
which indicates how similar the two objects are. 1 means equivalent and 0 means no similarity. An example of such a 
similarity measure is the Dice coefficient. In this case we have

\begin{align}
f: \mathbb{F}_2^k \times \mathbb{F}_2^k &\rightarrow [0,1], \\
((x_0, x_1, .. x_k), (y_0, y_1, .. y_k)) &\rightarrow \frac{2\sum_{i = 0}^k x_i y_i}{\sum_{i = 0}^k (x_i + y_i)}.
\end{align}
Note that the additions here are not denoted by $\oplus$ but the regular $+$ so they are not evaluated mod $2$. 

In order implement this advanced similarity tagging, we have to utilize the method of Gray synthesis which will be laid out 
in more detail in the coming section \ref{Gray_Synthesis}. All we need to know about it here is that it can synthesize 
arbitrary diagonal (in the computational basis) unitary matrices: For a given $2^k$ 
tuple of real numbers $\phi = (\phi_0, \phi_1, ..\phi_{2^k-1})$ a circuit $U_{\text{gray}}(\phi)$ can be synthesized such that for any computational 
basis state $\ket{y}$
\begin{align}
U_{\text{gray}}(\phi) \ket{y} = \text{exp}(i\phi_y)\ket{y}.
\end{align}
Note that Gray synthesis requires only up to $2^k$ CNOT gates, implying such a similarity tag is computationally 
more efficient or of equivalent efficiency as tagging with a multi-controlled Z gate.

We next show how applying Gray synthesis on the label register can be used to implement a tag, which acts as described in eq. 
\ref{similarity_phase_tag_eq}. Suppose we are given a similarity function $f$ of the type in eq. \ref{similarity_measure_eq}. 
Then the similarity tag $T^{\text{sim}}_f(q)$ for the query object  $q \in Q$ is given as:

\begin{align}
T^{\text{sim}}_f(q) &= U_{\text{gray}}(\phi_f^{\text{sim}}(q)),\\
\text{where\ \ } (\phi_f^{\text{sim}}(q))_y &= (-1)^y \pi f(q,y).
\end{align}

Next, we elucidate on the suitability of this operation as a similarity tag. For this we apply the similarity oracle 
consisting of the database encoding circuit\footnote{Note that this circuit doesn't necessarily has to encode a database but 
any other function calculating a binary value is also possible. One such alternative use case are optimization problems, which 
is under active research by the authors. Because of this we write $U_D\ket{x}\ket{0} = \ket{x}\ket{y(x)}$ instead of the 
$l(e(i))$ language used in the previous sections.} $U_D$ and the similarity tag to the index register in a uniform 
superposition.

\begin{equation}
	\label{similarity_oracle_action_eq}
	\begin{aligned}
	&O^\text{sim}(f,q,D) \ket{s}\ket{0}\\
	=&\sum_{x = 0}^{2^n -1} U_D^\dagger T^{\text{sim}}_f(q) U_D \ket{x}\ket{0} \\
	=&\sum_{x = 0}^{2^n -1} U^\dagger T^{\text{sim}}_f(q) \ket{x}\ket{y(x)}\\
	=&\sum_{x = 0}^{2^n -1} U^\dagger \text{exp}(i(-1)^{y(x)} \pi f(q,y(x)))\ \ket{x}\ket{y(x)}\\
	=&\sum_{x = 0}^{2^n -1} \text{exp}(i(-1)^{y(x)} \pi f(q,y(x))) \ket{x}\ket{0}.
	\end{aligned}
\end{equation}

We see that this oracle has the effect we assumed in eq. \ref{similarity_phase_tag_eq} with 
$\phi(x) = (-1)^{y(x)} \pi f(q,y(x))$. First of all, we note that (under reasonable assumptions about $f$) the alternating 
signs of the phases help getting $\phi_{\text{cm}}$ close to zero. 
To see why this is the case, we assume that for the majority of $x<2^n$, the statement $f(q,y(x))<0.5$ holds or in other 
words: Only in a few cases we actually do have similarity. Looking at the definition of $\phi_{\text{cm}}$ eq. \ref{cm_eq}, we 
observe the repeated addition of complex numbers mostly in the right half of the complex plane, which implies that the center 
of mass is most likely going to be in the right half of the complex plane. As the signs of the phases are alternating, 
we "balance" out the complex part by adding conjugates and non-conjugates which then 
implies $\phi_{\text{cm}} \approx 0$.
If we now look at eq. \ref{amplification_eq}, we see that the relation between the similarity measure $f(q,y(x))$ and 
the amplification factor $A_x$ is monotonical, which results in the desired behavior. Note that this doesn't imply that the 
amplification factor of $x$ is proportional to the value of $f$. Only the ordering is preserved, i.e.
\begin{align}
\text{if\ }&f(q, y(x_1)) < f(q, y(x_2))\\
\text{then\ }& A_{x_1} < A_{x_2}.
\end{align}

An example application of Grover's algorithm to a similarity oracle implementing the Dice coefficient can be found in 
fig. \ref{advanced_sim_search_plot}, where the above described aspects are clearly visible.

\subsection{Contrast Functions}
\begin{figure*}
	\begin{subfigure}{.5\textwidth}
		\includegraphics[width = \textwidth]{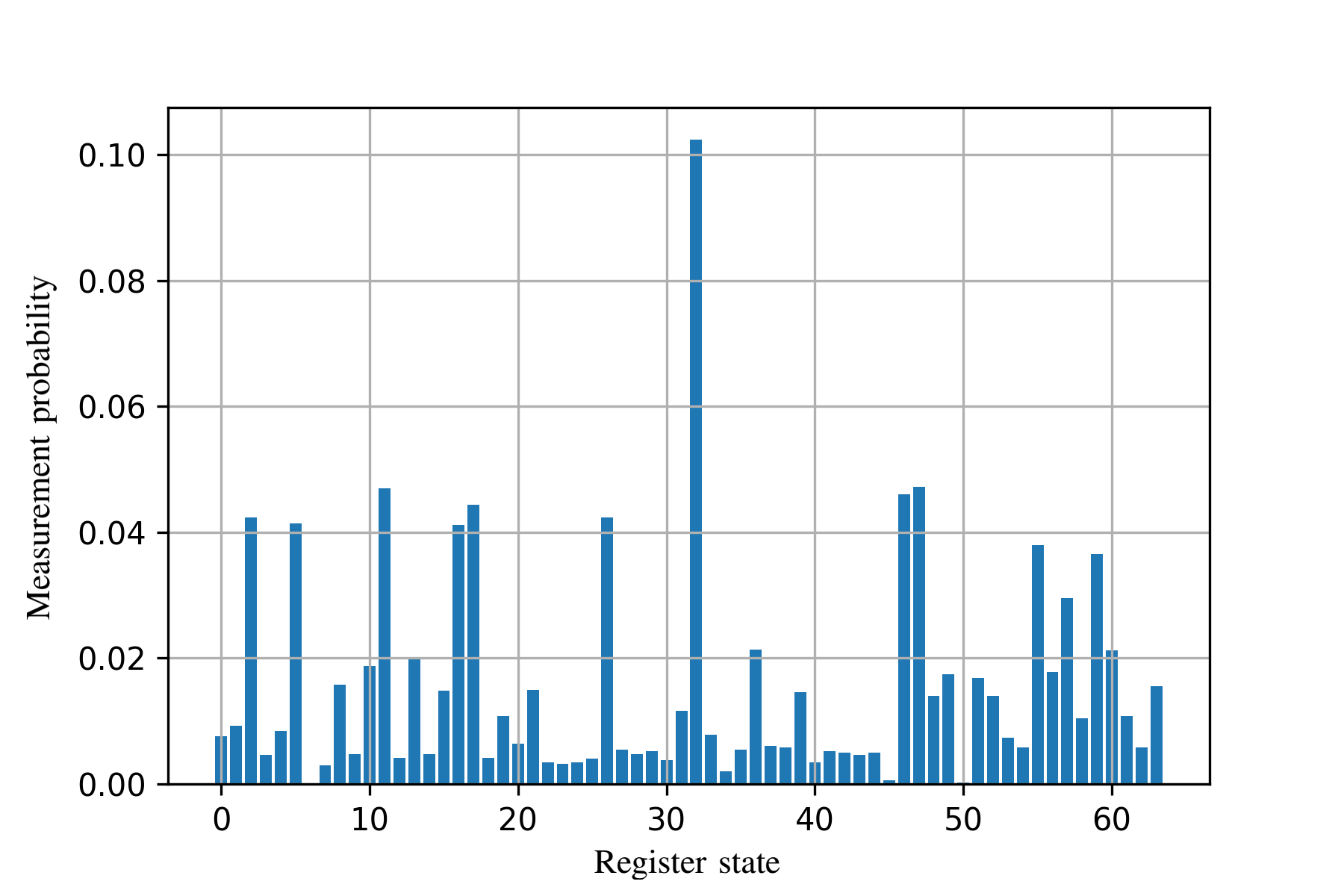}
			\caption{\label{advanced_sim_search_plot}}
	\end{subfigure}
	\begin{subfigure}{.5\textwidth}
		\includegraphics[width = \textwidth]{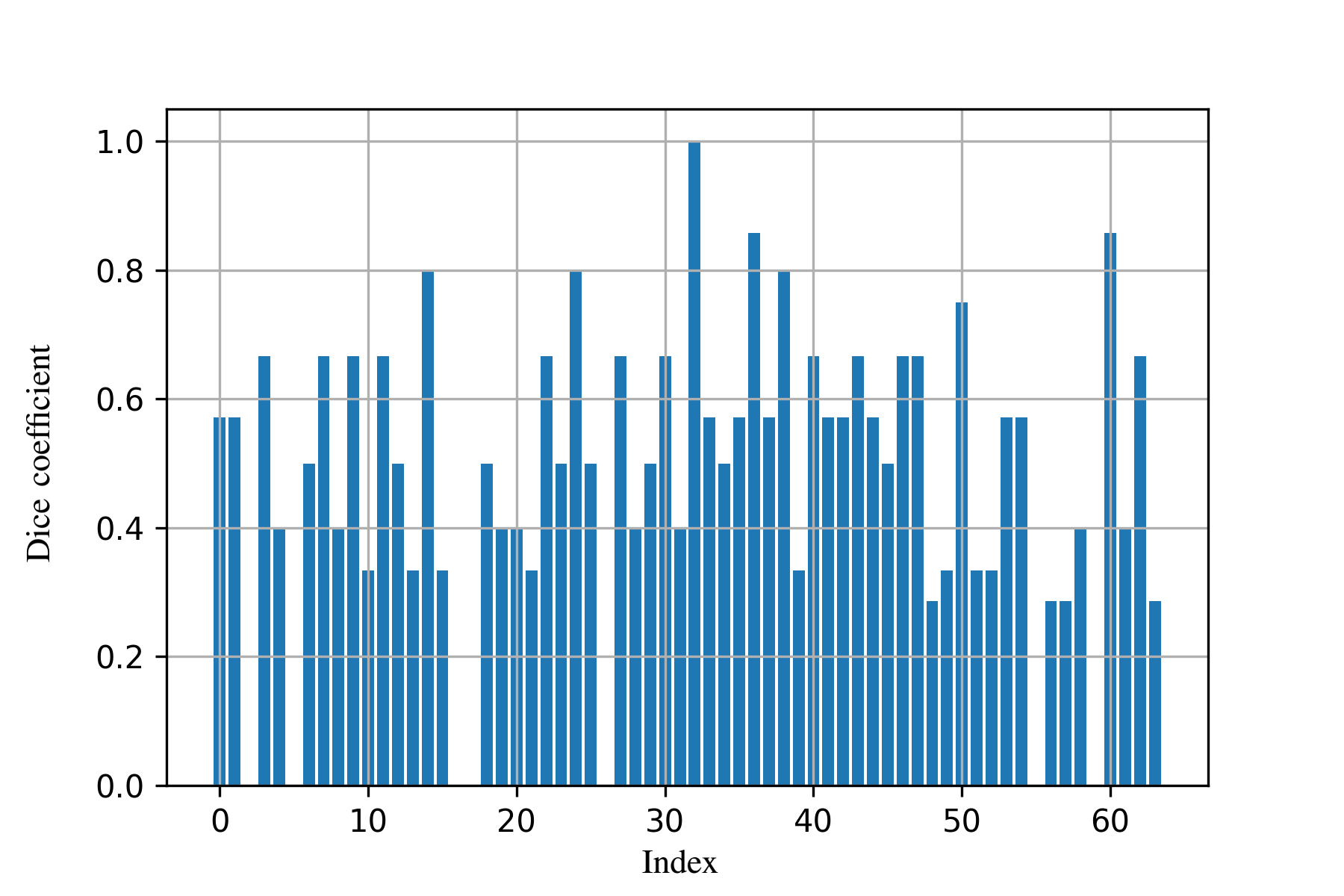}
		\caption{\label{dice_coeff_plot}}
	\end{subfigure}
	\begin{subfigure}{.5\textwidth}
		\includegraphics[width = \textwidth]{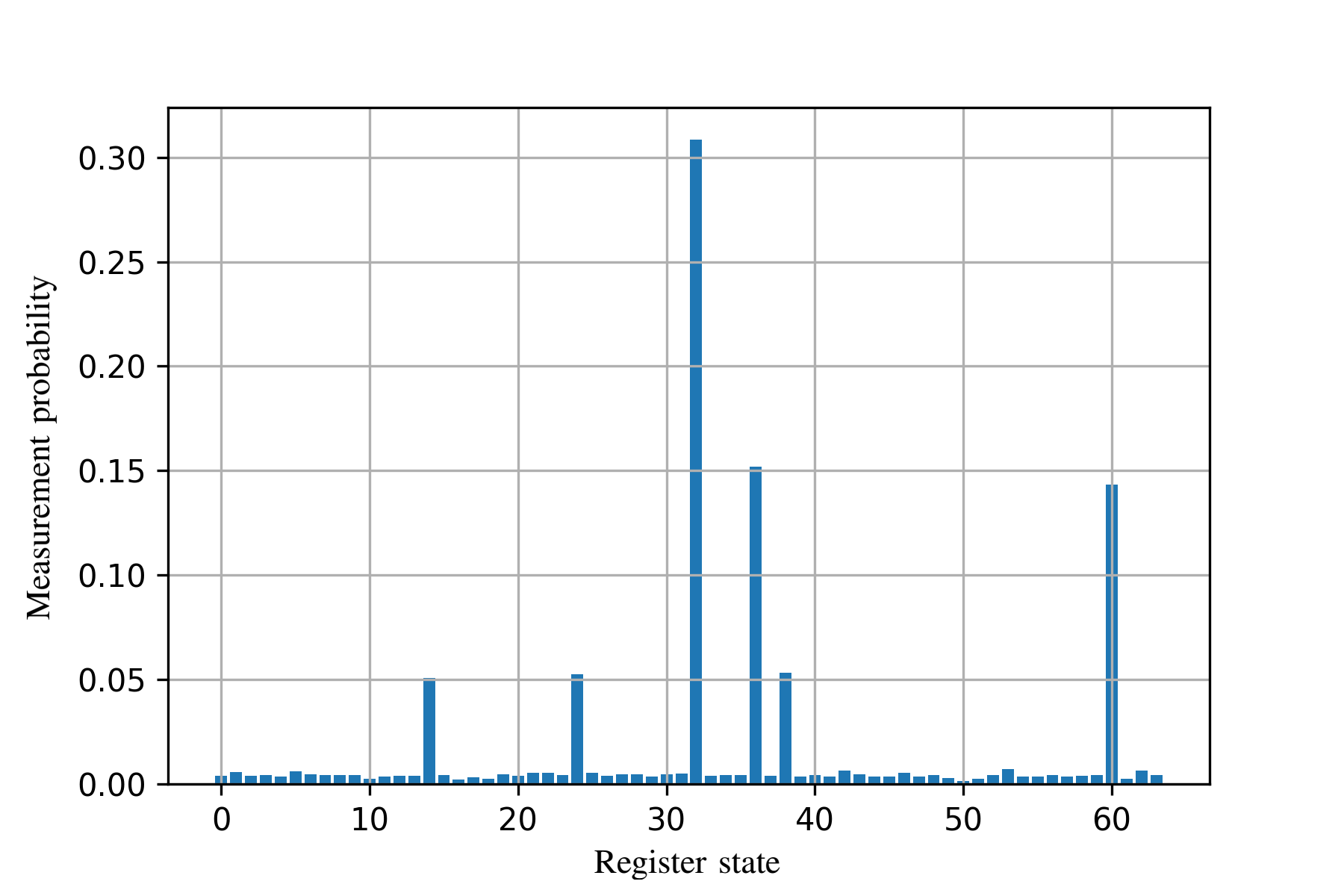}
		\caption{\label{advanced_sim_search_contrast_plot}}
	\end{subfigure}
	\begin{subfigure}{.5\textwidth}
		\includegraphics[width = \textwidth]{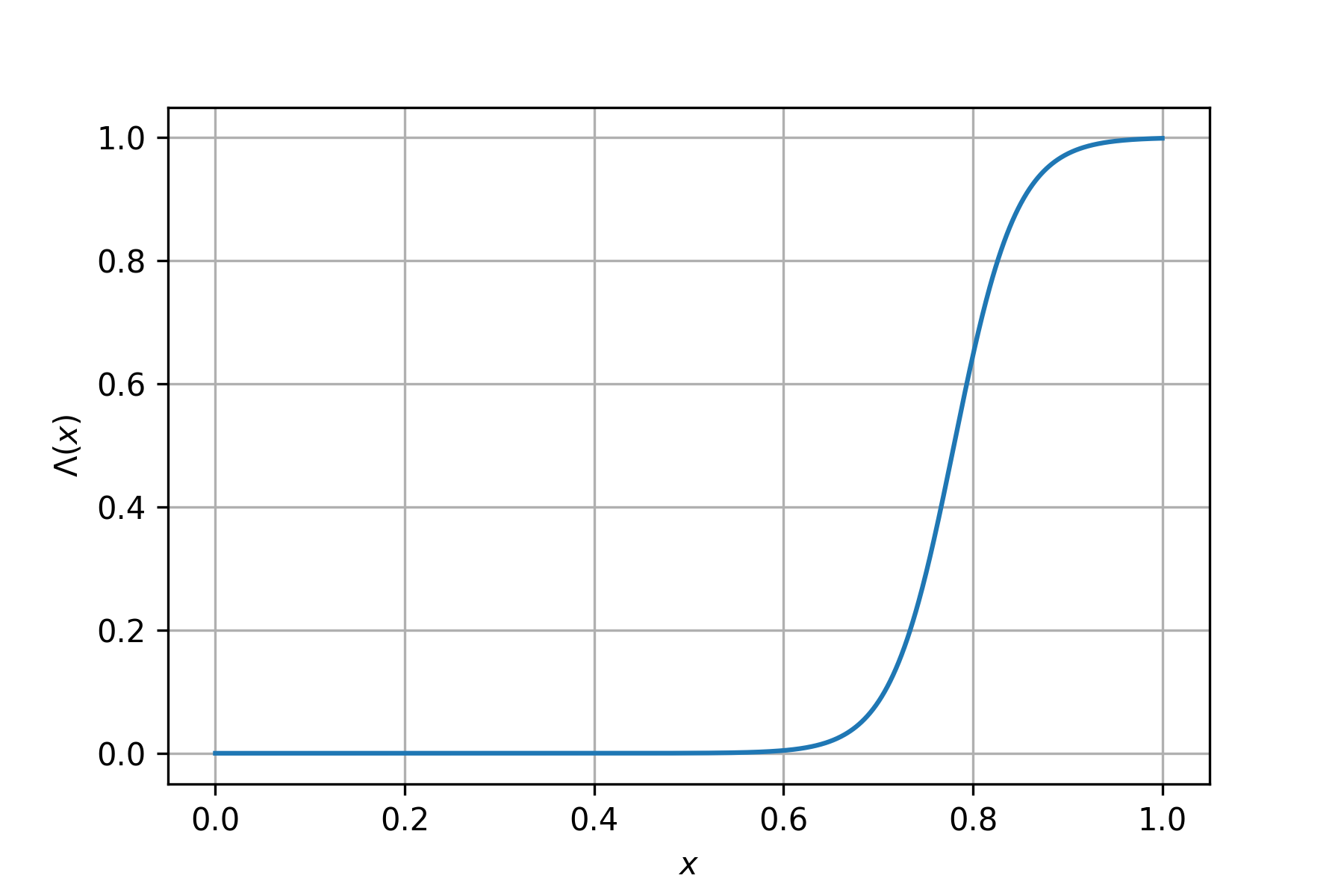}
		\caption{\label{contrast_function_plot}}
\end{subfigure}

\caption{\textbf{\ref{advanced_sim_search_plot}:} Plot of the measurement probability of the index register after application of Grover's algorithm to a similarity oracle using the Dice coefficient as similarity measure of an example database of 64 entries. \textbf{\ref{dice_coeff_plot}:} The similarity measure (Dice coefficient) used in the oracle from fig. \ref{advanced_sim_search_plot}. \textbf{\ref{advanced_sim_search_contrast_plot}:} The same plot as fig. \ref{advanced_sim_search_plot} but the similarity function has been wrapped in a contrast function. \textbf{\ref{contrast_function_plot}:} The contrast function used in fig. \ref{advanced_sim_search_contrast_plot}. The mathematical expression is $\Lambda(x) = (\text{exp}(30(0.78-x)) + 1)^{-1}$}
\end{figure*}

Since we are only interested in the ordering of the values of the similarity measure, we can improve some properties without changing information by applying a monotonically increasing function with signature
\begin{align}
\Lambda: [0,1] \rightarrow [0,1].
\end{align}
We call this a \textit{contrast function}. In other words: For a given similarity measure $f$ and a 
contrast function $\Lambda$, instead of 
$f$ we use $\tilde{f} = \Lambda \circ f$ as similarity measure 
(the $\circ$ denotes the composition\footnote{For two functions $g,h$ of 
fitting signature, the composition $g \circ h$ is the function which executes a successive application, i.e. 
$(g \circ h)(x) = g(h(x))$.}). An example of a contrast function which improved the results in many cases can be found 
in fig. \ref{contrast_function_plot}. Note that a large 
portion of the domain gets mapped to a value close to 0. This not only ensures the assumption $\tilde{f}(q,y(x))<0.5$ for 
most $x<2^n$ (required for $\phi_{\text{cm}} \approx 0$) but also yields $r_{\text{cm}} \approx 1$ as in most cases 
$\phi(x) \approx 0$ (compare eq. \ref{cm_eq}). This 
in turn gives us a good amplification factor $A_x$ for states that are supposed to be amplified but $A_x \approx 0$ for 
states that have only mediocre similarity (compare eq. \ref{amplification_eq}).
\\
\\
\\
\\

\label{sec:synthesis}
\section{Quantum Logic Synthesis}
\label{synthesis}

As pointed out in the previous sections, an integral part of generating oracles is the logic synthesis. 
There is a multitude of approaches each having their benefits and drawbacks. In this section, we focus on the 
\textit{Reed-Muller Expansion} and \textit{Gray Synthesis}. Subsequently, a new synthesis method developed by us is introduced, which is significantly more efficient in terms of general gate count. Our approach is to relax the constraint of all outputs having the same phase. This does not interfere with the outcome of Grover's algorithm as these phases cancel out during uncomputation. 
Finally, we derive and introduce another synthesis method, which addresses the pitfalls and requirements for scalable implementations of the belonging quantum circuits.

\subsection{Reed-Muller Expansion}
\label{rm_expansion}

The very first and foremost approach to tackle the challenge of logic synthesis is the Reed-Muller Expansion. 
Even though it is by far not the most efficient way to synthesize circuits for a quantum computer, 
it's basic concepts will be helpful for further understanding of the proposed concepts. 
The belonging method is denoted as Positive Polarity Reed-Muller expansion synthesis \cite{Porwik2002EFFICIENTCO}
and is abbreviated as PPRM accordingly for the rest of the paper.

First, we note that due to their reversible architecture, quantum computers do not have direct 
access to the full range of tools as in traditional 
logic synthesis. For example, we can't infer the input constellation of a classical AND gate by just looking at the output.
However, reversibility is a fundamental requirement to any quantum operation as there are only reversible building blocks available to construct said operation.
But there is a way to turn any non-reversible gate into a reversible one which is keeping the inputs in place and saving the result into a 
new qubit. Using an $n$-controlled $X$ gate we can therefore compute a multi-AND-gate between $n$-qubits into a new qubit. Acting with 
another (multi)-controlled $X$ gate on the same qubit, we can realize an XOR gate. Expressions of the type
\begin{align}
\label{xag_example}
(x_0 \text{ AND } x_1 \text{ AND } x_2) \text{ XOR } (x_0 \text{ AND } x_1)
\end{align}
are called XAG (\textbf{X}OR and \textbf{A}ND \textbf{G}raphs) and can be described very conveniently by polynomials over the boolean algebra $\mathbb{F}_2$. The corresponding polynomial for eq. \ref{xag_example} would be
\begin{align}
\label{xag_example_poly}
p(x) = x_0 x_1 x_2 \oplus x_0 x_1
\end{align}
Using this, we can introduce a very basic approach to quantum logic synthesis: Given a single column truth table $T$ depending on $n$ variables, its Reed-Muller expansion $RM_T^n(x)$ is the polynomial which is recursively generated by the following equation
\begin{align}
\label{RM-expansion}
RM_T^n(x) = x_0 RM_{T_1}^{n-1} (x) \oplus (x_0 \oplus 1) RM_{T_0}^{n-1} (x)
 \end{align}
Here $T_0,\, T_1$ denote the co-factors of $T$, i.e. the truth tables considering the entries of $T$ where $x_0 = 0$ or $x_0 = 1$. This recursion cancels at $n =0 $ with $RM^0_{\tilde{T}}(x)$ either equal to $0$ or $1$ depending on the value of $\tilde{T}$. The resulting polynomial is unique for any given truth table \cite{Kebschull1992}.
\begin{table}
\centering
\begin{tabular}{cc|ccc}
$x_0$ & $x_1$ & $T$ & $T_0$ & $T_1$\\
\hline
0 & 0 & 0 & 0 & -\\
0 & 1 & 1 & 1 & -\\
1 & 0 & 1 & - & 1\\
1 & 1 & 1 & - & 1\\
\end{tabular}
\caption{Example decomposition into co-factors. Note that while $T$ is depending on two boolean q variables, $T_0$ and $T_1$ are only depending on one variable.}
\end{table}
Given such a polynomial, one can generate the corresponding circuit implementation for a truth table by setting up a multi-controlled X gate for every "summand" of the polynomial.
\begin{figure*}
\begin{subfigure}[b]{.18\textwidth}
\begin{quantikz}
	\lstick{$\ket{x_0}$} & \ctrl{3} & \ctrl{3} & \qw \\ 
	\lstick{$\ket{x_1}$} & \ctrl{2} & \ctrl{2} & \qw \\ 
	\lstick{$\ket{x_2}$} & \ctrl{1} & \qw      & \qw \\ 
	\lstick{$\ket{t_0}$} & \targ{}  & \targ{}  & \qw
\end{quantikz}
\caption{\label{circ_xag_example}}
\end{subfigure}
\begin{subfigure}[b]{.33\textwidth}
\begin{quantikz}
	\lstick{$\ket{x_0}$} & \ctrl{2} & \qw      & \ctrl{2} & \qw      & \ctrl{1} & \ctrl{1} \\ 
	\lstick{$\ket{x_1}$} & \qw      & \ctrl{1} & \qw      & \ctrl{1} & \targ{}  & \targ{} \\ 
	\lstick{$\ket{x_2}$} & \targ{}  & \targ{}  & \targ{}  & \targ{}  & \qw 			& \qw
\end{quantikz}
\caption{\label{3_bit_parity_network}}
\end{subfigure}
\begin{subfigure}[b]{.45\textwidth}
\begin{quantikz}[column sep=0.2cm]
	\lstick{$\ket{x_0}$} & \ctrl{2} & \qw               & \qw      & \qw               & \ctrl{2} & \qw      & \ctrl{1} & \ctrl{1} & \qw & \qw \\ 
	\lstick{$\ket{x_1}$} & \qw      & \qw     				  & \ctrl{1} & \qw							 & \qw      & \ctrl{1} & \targ{}  & \targ{} & \gate{RZ(\frac{\pi}{2})} & \qw \\ 
	\lstick{$\ket{x_2}$} & \targ{}  & \gate{RZ(\frac{-\pi}{2})} & \targ{}  & \gate{RZ(\frac{-\pi}{2})} & \targ{}  & \targ{}  & \qw 	& \qw & \qw & \qw
\end{quantikz}
\caption{\label{circ_phi_synthesis}}
\end{subfigure}
\caption{\textbf{\ref{circ_xag_example}}: Example synthesis of the polynomial in eq. \ref{xag_example}. \textbf{\ref{3_bit_parity_network}}: The corresponding circuit of the parity matrix given in eq. \ref{parity_matrix}. The sequence of parity operators traversed is $ x_0 \oplus x_2, x_0 \oplus x_1 \oplus x_2, x_1 \oplus x_2, x_2, x_0 \oplus x_1, x_1, x_0, \emptyset$.  \textbf{\ref{circ_phi_synthesis}}: Example synthesis of $\phi = \pi (1,0,1,1,0,1,1,1,0)$. The solution of eq. \ref{phase_equation} is $\theta = \frac{\pi}{2} (-1, -1, 0, 0, 0, 1, 0, 2)$. The phenomenon of the last 2 CNOTs canceling out is treated in section \ref{parity_traversal}}
\end{figure*}
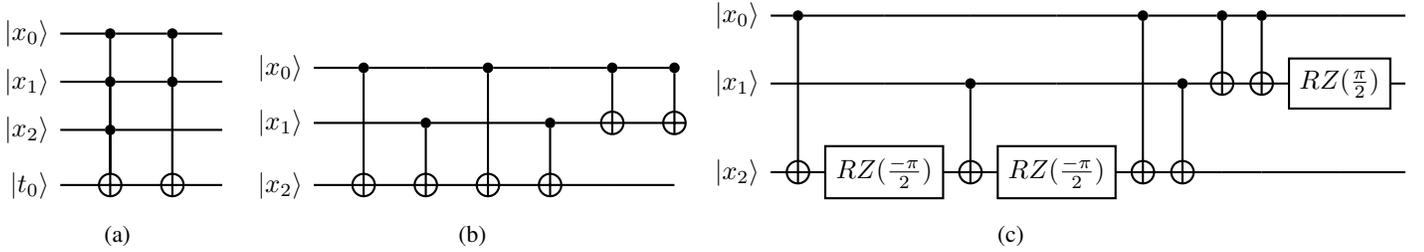
While in principle sufficient, this method still has a lot of room for improvement as multi-controlled gates are computationally very 
expensive - \cite{Abdollahi2013} an $n$-controlled NOT gate requires $(2n^2-2n+1)$ CNOT gates. For this 
reason, the following sections investigate further methods towards a more resource efficient synthesis of boolean functions.

\subsection{Gray Synthesis}
\label{Gray_Synthesis}
A more efficient synthesis method can be achieved by taking advantage of the non-classical properties of a quantum computer. 
This method is called Gray Synthesis and a detailed description is given in \cite{Amy2018}. The method works with three types of gates: CNOT, H and $\text{T}_m$. While \cite{Amy2018} also considers the possibility of directly synthesizing logic functions, these functions have to be $\mathbb{F}_2$-linear, which is an impermissible restriction in our case of application.  In this paper, the focus on the possibility to synthesize a user-determined phaseshift for each computational basis state (now called "input state").
This in turn can then be used to efficiently synthesize a truth table $T(x)$ by applying the method to the input register combined with 
the output qubit that is enclosed with $H$-gates. To illustrate how this works: consider that we synthesize a phase-shift of $0$ for every 
input state that has a $0$ in the output qubit and a phase-shift of $\pi T(x)$ for every state that has a $1$ in the output qubit. This 
leads to the following calculation towards the synthesis of the desired truth table:
\begin{equation}
\begin{aligned}
&H_{\text{out}}U_{\text{gray}}H_{\text{out}}\ket{x}\ket{0}\\
= &\frac{1}{\sqrt{2}}H_{\text{out}}U_{\text{gray}}\ket{x}(\ket{0} + \ket{1})\\
=&\frac{1}{\sqrt{2}}H_{\text{out}}\ket{x}(\ket{0} + \text{exp}(i\pi T(x)) \ket{1})\\
\label{phase_differences}= &\frac{1}{\sqrt{2}}H_{\text{out}}\ket{x}(\ket{0} + (-1)^{T(x)} \ket{1})\\
= &\ket{x}\ket{T(x)}.
\end{aligned}
\end{equation} \\
Here, the first ket $\ket{x}$ represents the input register, the second ket $\ket{0}$ the output qubit, and $U_{\text{gray}}$ is the 
circuit which synthesizes the desired phases.

Since a good understanding of the phase synthesis method is a prerequisite to fully capture the power of phase tolerant synthesis, we will 
now go on to give a summary on how the phases are synthesized. A key concept in this context are parity operators, which basically are XOR expressions of different combinations of input variables e.g. $x_0 \oplus x_2 \oplus x_3$. Keep in mind, that we denote the XOR gates with the $\oplus$ symbol, because additions over real numbers will occur now too. The value of a parity operator on a given input state can be "loaded" into a qubit by applying a sequence of CNOT gates (see fig. \ref{3_bit_parity_network} for an 
example). We can use the notion of parity networks to assign the desired phase to the input state by applying RZ gates onto the parity 
operators. Since RZ gates have the matrix representation 
$\text{diag}(\text{exp}\left(\frac{-i\theta}{2}\right), \text{exp}\left(\frac{i\theta}{2}\right))$, the change of phase $\Delta \phi $ after applying the gate RZ$(\theta_p)$ to a qubit, which has the parity operator $p$ loaded can be written as
\begin{equation}
\begin{aligned}
\Delta \phi = \begin{cases} - \frac{\theta_p}{2} & \text{if } p(x) = 0\\ + \frac{\theta_p}{2} & \text{if }  p(x) = 1 \end{cases}\\
= \frac{1}{2} (-1)^{p(x)} \theta_p.
\end{aligned}
\end{equation}

 the phase applied on a given input state $x = (x_0, x_1, .. x_n)$ after traversing the set of parity operators $P$ can thus be summarized as
\begin{align}
\phi_x = \frac{1}{2} \sum_{p \in P} (-1)^{p(x)}\theta_{p}.
\end{align}
Note that this is not a sum of $\mathbb{F}_2$ elements but of real numbers. We can therefore control what kind of phase each input state receives 
by carefully deciding on how to distribute the phase shifts $\theta_{p}$. As each input state $x$ can be uniquely identified\footnote{To 
see that this is true, consider two input states, where $i$ is an index where the states differ. The parity operator $x_i$ thus has 
differing values on the two states.} by just looking at the set of parity operators, which return $1$ when applied to $x$, we can give each 
state a unique constellation of phase shifts by iterating over every
possible parity operator. The next question is, how to determine the required 
$\theta =(\theta_{x_0}, \theta_{x_1}, \theta_{x_0 \oplus x_1}...)$ phase shifts for a given sequence of desired overall phase shifts 
$\phi = (\phi_0, \phi_1 .. \phi_{2^n})$? In this regard, by successively writing down, which state receives which phase shift one 
can set up a system of linear equations. The resulting matrix is denoted as the \textit{parity matrix} $D$. The corresponding 
system of equations therefore yields
\begin{align}
\label{phase_equation}
\phi = \frac{1}{2} D \theta.
\end{align}
This is achieved by ordering the rows according to the natural order of the input states (ie. $000,001,010$..) and the columns according to an algorithmic solution of the Hamming TSP\footnote{The Hamming TSP stands for a TSP 
(Travelling Salesman Problem) instance, where the Hamming distance is used as a distance measures between the involved entities/objects.}, which visits all parity operators. The resulting matrix only depends on the number of input qubits. In the case of 3 input qubits we get
\begin{align}
\label{parity_matrix}
D_3 = \begin{pmatrix}1 & 1 & 1 & 1 & 1 & 1 & 1 & 1\\-1 & -1 & 1 & 1 & -1 & 1 & -1 & 1\\1 & -1 & -1 & 1 & -1 & -1 & 1 & 1\\-1 & 1 & -1 & 1 & 1 & -1 & -1 & 1\\-1 & -1 & -1 & -1 & 1 & 1 & 1 & 1\\1 & 1 & -1 & -1 & -1 & 1 & -1 & 1\\-1 & 1 & 1 & -1 & -1 & -1 & 1 & 1\\1 & -1 & 1 & -1 & 1 & -1 & -1 & 1\end{pmatrix}
\end{align}
Note that the last column corresponds to the "empty" parity operator, which we define to be $0$ on every input state. 
As the corresponding coefficient $\theta_{\emptyset}$ only induces an irrelevant global phase, it can be ignored during synthesis.
Solving systems of linear equations is possible in $\mathcal{O}(N^3)$. Even though we only have to perform the synthesis once per database, this would still scale very bad compared to classical searching. A much more efficient solution is based on the following considerations: According to \cite{Meuli2020} the $i$-th row of the Hadamard-Matrix $H_N$ of degree $N$ is the truth table of parity operator $i$. To be more precise the $x$-th row of the $i$-th column is equal to
\begin{align}
(H_N)_{xi} &= (-1)^{i(x)},\\
\text{where }i(x) &= \bigoplus_{k = 0}^{n-1} i_k x_k.
\end{align}
Since $H_N$ is symmetric, this applies to the columns 
as well. Therefore our $D$ is simply the Hadamard-Matrix with permuted columns, because this is precisely how the columns of $D$ are defined.
\begin{align}
H_N = D \Sigma,
\end{align}
where $\Sigma$ is some permutation matrix.
We can now use a well know property of $H_N$:
\begin{equation}
\begin{aligned}
&H_N H_N^t = N \mathbb{I}\\
\Leftrightarrow &H_N^{-1} = \frac{H_N^t}{N} = \frac{H_N}{N},
\end{aligned}
\end{equation}
to find the inverse of $D$
\begin{equation}
\begin{aligned}
D^{-1} &= (H_N \Sigma^{-1})^{-1}\\
&= \Sigma H_N^{-1}\\
&= \frac{\Sigma H_N}{N}.
\end{aligned}
\end{equation}
Applying this to eq. \ref{phase_equation}, we get
\begin{equation}
\begin{aligned}
\label{hw_transform_eq}
\theta = 2D^{-1} \phi = \frac{2\Sigma H_n \phi}{N}.
\end{aligned}
\end{equation}
This can be interpreted as follows: $\theta$ is the Hadamard-Walsh transform\footnote{The Hadamard-Walsh transform
\cite{Walsh_Hadamard} is closely related to Fourier analysis. 
It transforms complex numbers to a spectrum of orthogonal functions - these functions are called Walsh 
functions.} of $\phi$ up to some permutation and factor. Fortunately, 
$H_n$ doesn't need to be explicitly calculated, since there is an algorithm called \textit{Fast Hadamard Walsh Transformation}
which performs the transformation in $\mathcal{O}(\text{log}(N)N)$ \cite{Fino1976}.

\subsection{Parity Operator Traversal}
\label{parity_traversal}

As the space of parity operators is the dual space for the vector space $\mathbb{F}_2^n$, it can be indexed by the natural 
numbers $<2^n$. Finding a parity operator traversal route therefore reduces to finding a solution to an integer 
programming \footnote{The Hamming TSP and the integer/binary programming TSP are closely related, given that 
within the Hamming TSP we have binary strings as the entities and the Hamming distance as the distance 
between those entities. In this regard, the integer programming constitutes the problem formulation as a combinatorial 
problem with an optimization function over integers/bitstrings.} traveling salesman problem \cite{INT_TSP_1960}. 
The restricting feature here is, that we need one CNOT gate 
per bit that is changed in a traversal step. A possible solution for this is the Gray-Code \cite{Bhat1996BalancedGC}. The Gray-Code is a sequence of integers which traverse every single natural number $<2^n$ by only changing a single bit in every 
step. Even though it seems like this is precisely what is required, there is one drawback. As the Gray-Code traverses 
\textbf{every} parity operator, there is a considerable overhead, since only parity operators, which 
have a non-vanishing Hadamard-coefficient need to be traversed. This problem is addressed by a very simple heuristic 
solution. In order to ensure our traversal route ends where it started, we introduce a second salesman, which will meet the 
first after they both traversed the required integers. The resulting route will then be the route 
of the first salesman concatenated with the reversed route of the second salesman. Regarding the behavior of the individual 
salesmen, they always choose the closest parity operator, which has not been visited by either of them. We tried penalizing them for moving very far apart in order to reduce the reunification cost, which however didn't lead to a significant reduction of the overall path length. Using this technique yields a (scale dependent) cut of about 10\% in traversed parity operators, however it is 
significantly more expensive regarding the classical resources.

\subsection{Phase Tolerant Synthesis}

Even though Gray synthesis is already significantly cheaper in terms of CNOT gates than PPRM synthesis,
there is an even bigger possible optimization with regard to the
required quantum resources. During oracle application, we observe, that the label register always needs to be  uncomputed after the winner state is tagged. This implies that we can be \textit{tolerant} regarding the phases of the label 
variable as the tagging gate is not concerned with the involuntarily synthesized "garbage phases". To be more explicit, the synthesis methods discussed so far applied to an index register in uniform superposition 
would result in the state
\begin{align}
\sum_{x = 0}^{2^n-1} \ket{x}\ket{T(x)}.
\end{align}
However a state like
\begin{align}
\sum_{x = 0}^{2^n-1} \text{exp}(i\chi_x)\ket{x}\ket{T(x)},
\end{align}
will be tagged in the correct way too and the "garbage-phases" $\chi_x$ will be uncomputed, when the label register 
ism uncomputed. In order to understand how to synthesize with garbage-phases, we note that in this method of synthesis, the 
logical information is only captured in the phase \textit{differences} of the $0$ and $1$ state of the output qubit. 
Following the principles applied on eq. \ref{phase_differences}, we arrive at:
\begin{equation}
\begin{aligned}
&H(\text{exp}(i\phi_0) \ket{0} + \text{exp}(i\phi_1) \ket{1})\\
& =\text{ exp}(i\phi_0)H( \ket{0} + \text{exp}(i(\phi_1-\phi_0)) \ket{1})\\
=& \begin{cases}
 \text{exp}(i\phi_0) \ket{0}  \text{ if } \phi_1-\phi_0 = 0\\
 \text{exp}(i\phi_0) \ket{1}  \text{ if } \phi_1-\phi_0 = \pi.
 \end{cases}
\end{aligned}
\end{equation}
This observation can now be used in combination with a profitable choice of sequence of parity operators: Note how the parity network 
given in fig. \ref{3_bit_parity_network} visits every parity operator, which includes $x_2$ first and after that only operators which do 
not contain $x_2$. Furthermore directly after every parity operator which includes $x_2$ is visited, every circuit wire 
contains it's very own parity operator, i.e. wire $0$ contains the parity operator $x_0$, wire $1$ contains $x_1$ and so on. Therefore, we 
can simply end the synthesis procedure at this point, because every bit of phase that is synthesized after, doesn't differentiate between 
the $\ket{0}$ and the $\ket{1}$ state of the output qubit (which would be $x_2$ here), and therefore doesn't contribute anything to the phase 
difference of the $\ket{0}$ and 
the $\ket{1}$ state.\\
To make this notion accessible from a more formal point of view, consider the case that we are applying Gray synthesis to the state of uniform superposition $\ket{s}$:
\begin{align}
\ket{s} = \frac{1}{\sqrt{2^n}}  \sum_{x = 0}^{2^n} \ket{x}.
\end{align}
We factor out the output qubit
\begin{align}
\ket{s} = \frac{1}{\sqrt{2^n}}  \sum_{x = 0}^{2^{n-1}} (\ket{0} + \ket{1})\ket{x}.
\end{align}
The "first step" in this view of Gray synthesis is synthesizing phases $\phi_{(0,x)}$ and $\phi_{(1,x)}$ on the output qubit
\begin{align}
\label{first_step_eq}
&U_{\text{gray}}^{\text{Step 1}} \ket{s} =\\
&\frac{1}{\sqrt{2^n}} \sum_{x = 0}^{2^{n-1}} \left(\text{exp}(i\phi_{(0,x)})\ket{0} + \text{exp}(i\phi_{(1,x)})\ket{1}\right)\ket{x} =: \ket{\psi} \notag.
\end{align}
The second step would now be to synthesize a phase on the remaining qubits which could be seen as "correcting" the garbage phase $\chi_x$. This however does not change the relative phase of the $\ket{0}$ and $\ket{1}$ state of the output qubit
\begin{align}
&U_{\text{gray}}^{\text{Step 2}} \ket{\psi} =\\
&\frac{1}{\sqrt{2^n}}  \sum_{x = 0}^{2^{n-1}} \text{exp}(-i\chi_x)\left(\text{exp}(i\phi_{(0,x)})\ket{0} + \text{exp}(i\phi_{(1,x)})\ket{1}\right)\ket{x}. \notag
\end{align}
The important point of phase tolerant synthesis is that the \textit{difference} $\phi_{(0,x)} - \phi_{(1,x)}$ determines the result of the logic output state:
\begin{align}
\phi_{(0,x)} - \phi_{(1,x)} = \begin{cases} 0 &\text{if } T(x) = 0 \\ \pi &\text{if } T(x) = 1. \end{cases}
\end{align}\\
Note that while we in principle could determine the phases $\chi_x, \phi_{(0,x)}, \phi_{(1,x)}$ from the user specified input phases $(\phi_0, \phi_1.. \phi_{2^n})$, it is neither necessary nor relevant. Just breaking the routine after the first step eq. \ref{first_step_eq} was performed, is sufficient.\\

\begin{figure*}
\begin{subfigure}[b]{.5\textwidth}
\begin{quantikz}[column sep=0.16cm]
	\lstick{$\ket{x_0}$} & \qw & \ctrl{2}     & \qw      & \qw               & \ctrl{2} & \qw      & \ctrl{1} & \qw & \ctrl{1} & \qw \\ [0.15cm]
	\lstick{$\ket{x_1}$} & \qw & \qw      	  & \ctrl{1} & \qw							 & \qw      & \ctrl{1} & \targ{}  & \gate{Rz\left(\frac{-\pi}{2}\right)} & \targ{} & \qw \\[-0.1cm]
	\lstick{$\ket{x_2}$} & \gate{H} & \targ{} & \targ{}  & \gate{Rz\left(\frac{\pi}{2}\right)} & \targ{}  & \targ{}  & \gate{Rz\left(\frac{\pi}{2}\right)} 	& \gate{H} & \qw & \qw
\end{quantikz}
\caption{\label{fig:circ_2input}}
\end{subfigure}
\begin{subfigure}[b]{.5\textwidth}
\centering
\begin{quantikz}[column sep=0.3cm, row sep = 0.6cm]
	\lstick{$\ket{x_0}$} & \qw & \ctrl{2}     & \qw      & \qw               & \ctrl{2} & \qw      & \qw  & \qw & \qw \\[0.3cm]
	\lstick{$\ket{x_1}$} & \qw & \qw      	  & \ctrl{1} & \qw							 & \qw      & \ctrl{1} & \qw  & \qw & \qw \\[0.1cm]
	\lstick{$\ket{x_2}$} & \gate{H} & \targ{} & \targ{}  & \gate{Rz\left(\frac{\pi}{2}\right)} & \targ{}  & \targ{}   & \gate{Rz\left(\frac{\pi}{2}\right)} 	& \gate{H} & \qw
\end{quantikz}
\caption{\label{fig:circ_2input_phasetolerant}}
\end{subfigure}
\caption{\textbf{\ref{fig:circ_2input}}: Logic synthesis of the 2 input-bit truth table 1001. \textbf{\ref{fig:circ_2input_phasetolerant}}: Phase tolerant 
logic synthesis of the 2 input-bit truth table 1001. Note that we only need to use 4 instead of 6 CNOT gates. The savings converge to 50\%
for larger truth tables.}
\end{figure*}
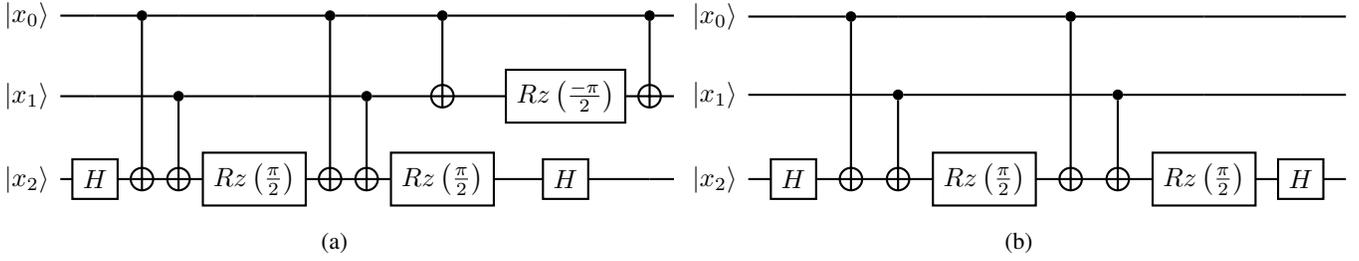

\section{CSE Synthesis}
\label{performance_optimization}

In this section, we present further progress beyond state-of-the-art from our research activities - the implementation
of quantum logic synthesis under the consideration of restrictions in a real world setup. 
Phase tolerant Gray synthesis may be very resource friendly to both classical and quantum resources, however scaling on 
real devices is still challenging: Taking a look at eq. \ref{hw_transform_eq} we see that, because the only values of the entries of $\phi$ that can appear in the scenario of logic synthesis are $\pm \frac{\pi}{2}$. Therefore the values of the entries of $\phi$ are integer multiples of $\pm \frac{\pi}{N}$. This implies that if we want to encode an array with $N$ entries, we will need reliant $\frac{\pi}{N}$ RZ gates\footnote{According to \cite{McKay_2017} it is possible to implement RZ gates virtually without any error and duration, however this might not be the case for every physical qubit realization} are. As this quantity will be central in the upcoming discussion, we will refer to the \textit{$T$-order} of a circuit by the minimal number 
$m\in \mathbb{N}$, such that every phase gate can be expressed as an integer multiple of a $\frac{\pi}{2^m}$ phase gate\footnote{According 
to \cite{Zeng2007} a logical qubit of a circuit of order $m$ in fault tolerant implementation (using the punctured Reed-Muller Code) requires (up to) $2^{m+2}-1$ physical qubits}.

Another ineffectiveness we observe is the fact that there is no usage of redundancy for the synthesis of multiple truth table columns even though we know that we will synthesize about $\mathcal{O}(\text{log}(N))$ columns. In the worst case we have the same 
column twice which means two syntheses procedures, although one synthesis + 1 CNOT gate would be sufficient. This handmade solution requires about half the resources, raising the question for an automatisation of this procedure.

Both of these problems can be tackled with by an approach based on calculating intermediate results
and intelligently optimizing the next synthesis steps accordingly.
Such an approach has already been proposed in \cite{Meuli2020}. Even though the authors could effectively demonstrate a 
reduction in the T-order, their technique still does not consider redundancies, since they focus on synthesizing truth 
tables with only a single column. Even though very heavy on the classical resources side and moderate on the qubit count, our 
approach (presented in this paper) has been shown to significantly reduce the T-order (see table \ref{Comparison_AVG_METRICS_CSE_SYNTH}). The basic idea lies in automated algebraic simplifications. As we saw in section \ref{rm_expansion}, a single column truth-table can be represented by a $\mathbb{F}_2$-polynomial. Multi-column truth tables can be therefore be interpreted as tuples of $\mathbb{F}_2$ polynomials $f(x) = (f_1(x), .. f_n(x))$. The key step is now to apply the common sub-expression elimination (CSE) algorithm of the computer algebra system (CAS) library 
\textit{sympy}\footnote{SymPy CAS: https://www.sympy.org, as of 03.08.2021} to these polynomials. This will give a sequence of intermediate 
values which reduce the overall resources required for the evaluation of $f$. For example, given the truth table

\begin{center}
\begin{tabular}{ccc|ccc}
$x_0$ & $x_1$ & $x_2$ & $f_0$ & $f_1$ & $f_2$\\
\hline
0 & 0 & 0 & 0 & 1 & 0\\
0 & 0 & 1 & 1 & 0 & 0\\
0 & 1 & 0 & 1 & 0 & 0\\
0 & 1 & 1 & 0 & 1 & 0\\
1 & 0 & 0 & 1 & 1 & 0\\
1 & 0 & 1 & 1 & 0 & 0\\
1 & 1 & 0 & 0 & 0 & 0\\
1 & 1 & 1 & 0 & 1 & 1\\
\end{tabular}
\end{center}
by applying Reed-Muller Expansion we get:
\begin{subequations}
\begin{align}
f_0(x) &= x_0 x_2 \oplus x_0 \oplus x_1 \oplus x_2\\
f_1(x) &= x_0 x_1 \oplus x_0 x_2 \oplus x_1 \oplus x_2 \oplus 1\\
\label{raw_polynomials}
f_2(x) &= x_0 x_1 x_2.
\end{align}
\end{subequations}
We now apply the CSE algorithm, which yields two intermediate values $g_0$ and $g_1$:
\begin{subequations}
\begin{align}
g_0(x) &= x_0 x_2\\
g_1(x) &= g_0(x) \oplus x_1 \oplus x_2.
\end{align}
\end{subequations}
The results utilizing the intermediate values are now:
\begin{subequations}
\begin{align}
\label{eliminated_polynomials}
f_0(x) &= g_1(x) \oplus x_0\\
f_1(x) &= g_1(x) \oplus x_0 x_1 \oplus 1\\
f_2(x) &= g_0(x) x_1.
\end{align}
\end{subequations}
Note that a product of $k$ variables will give a $k$-controlled X gate in PPRM synthesis\footnote{Note that basic PPRM
synthesis can also be supported by phase tolerant Gray synthesis by outsourcing the synthesis of the multi-controlled gates 
to the phase tolerant algorithm}. Synthesizing such a gate is equivalent to a truth table with $2^m$ entries, which implies 
that the T-order for a circuit containing only $m$-products or lower is $m$. Note that the synthesis of eq. 
\ref{raw_polynomials} contains a product of three variables, which implies that it's circuit has a T-order of 3, while the 
synthesis of eq. \ref{eliminated_polynomials} only contains products of order 2 implying a T-order of 2. We therefore 
successfully lowered the T-order by 1 at the cost of $\frac{2}{3}$ qubit overhead per truth table column. Larger synthesis yield much higher gains in the T-order but also 
much higher qubit overhead and additional CNOT gates when compared to "pure" phase tolerant Gray synthesis. However we reiterate that pure Gray synthesis is not scalable for arbitrary hardware architectures, demanding these drawbacks for real world applications.\\
Another point we would like to highlight is that the resulting sub expressions do not have to be synthesized with PPRM but an arbitrary synthesis method. In our implementation, we select the method of lowest CNOT count from a pool of methods.
This pool contained the Gray synthesis and PPRM synthesis implemented in tweedledum, as well as phase tolerant synthesis and a custom version of the PPRM synthesis, where multicontrolled X gates are being outsourced to a phase tolerant algorithm. For
each step of the array oracle synthesis, each method from the pool
of synthesis options is tested individually and the optimal method is
selected by means of the lowest CNOT count. It showed, that either
phase tolerant Gray synthesis or the custom phase tolerant supported
PPRM synthesis are optimal in many steps of the synthesis process.

While our implementation might not be directly feasible due to its high demand for classical resources, the general idea seems to yield improvements. We therefore leave the search of an advanced common subexpression elimination algorithm specifically tailored for $\mathbb{F}_2$-polynomials as an open research question. Regarding the overhead in quantum resources we note that it is always possible to resubstitute equations back into each other. As this increases the T-order for the expressions in question, it is not directly clear in which cases doing this is viable. Another open question therefore arises for an automatic procedure, which decides whether an intermediate value is worth calculating.

\label{sec:benchmarking}
\section{Benchmarking}
\label{benchmarking}

\subsection{Comparison of different synthesis methods}

\begin{table*}[t]
    \centering
    \begin{tabular}{lllllllllllllll}
        \toprule
        \multirow{2}{*}[-1em]{Data size} & \multicolumn{4}{c}{Average CNOT count} & \multicolumn{1}{c}{} & \multicolumn{3}{c}{Average $U$ count} &\multicolumn{1}{c}{} & \multicolumn{2}{c}{Average $\text{T}_m$ count (T-order)}& \multicolumn{1}{c}{} & \multicolumn{2}{c}{Qubits}\\
        \addlinespace[10pt]
        \cmidrule(lr){2-5} \cmidrule(lr){7-9} \cmidrule(lr){11-12} \cmidrule(lr){14-15} \\
        {} & PT (Gray-code) & PT (HTSP)& Gray & Difference & {} & PT & Gray & Difference & {} & PT & Gray & {} & PT & Gray\\
        \midrule
        4 & 8  & 6 & 6 & 0.0\% & &9 & 12 & 33.3 \% & & 6 (4) & 10 (4)& &4 & 4\\
        8 & 24 & 19 & 33 & 73.7\% & & 23 & 40 & 73.9\% & &23 (5) & 38 (5) & &6 & 6 \\
        16 & 64 & 59 & 103 & 74.6\% & & 60 & 110 & 83.3\% & &63 (6) & 119 (5)& &8 & 8 \\
        32 & 160 & 148 & 280 & 89.2\% & & 149 & 279 & 87.3\% & &189 (7) &356 (6)& &10 & 10  \\
        64 & 384 & 365 & 709 & 94.9\% & &363 & 693 & 90.9\% & &547 (8) & 1015 (8) &&12 & 12 \\
        128 & 896 & 861 & 1713 & 99.0\% & & 848 & 1684 & 98.6\% & &1399 (9) & 2804 (9) &&14 & 14 \\
        256 & 2048 & 1979 & 3967 & 100.5\% & & 1963 & 3908 & 99.1\% & &3636 (10) & 7132 (10) & &16 & 16  \\
        512 & 4608 & 4509 & 9022 & 100.1\% & & 4466 & 8920 & 99.7\% & &9019 (11)& 18225 (11) & &18 & 18  \\
        1024 & 10240 & 10116 & 20126 & 99.0\% & & 9985 & 19937 & 99.7\% & &22001 (12) & 44835 (12) &&20 & 20 \\
        \bottomrule
    \end{tabular}
    \caption{Comparison of average gate counts between our phase tolerant (PT) synthesis and the \textit{Tweedledum} implementation of Gray synthesis.}
		\label{Comparison_PTS_GRAY}
\end{table*}
In order to provide meaningful data about synthesis performance, we first elaborate on our benchmarking method. We compare our method for synthesizing Grover oracles to the \textit{Qiskit} implementation and algorithms from the synthesis library \textit{tweedledum} \cite{tweedledum}.\\
The evaluation task is to synthesize a database circuit $U_D$ eq. \ref{database_circuit} corresponding to an array of data which is randomly generated. The lengths of these arrays are discretely being increased in increments of powers of two, resulting in a range from 4 to 1024 data entries. We choose the label size $k$ equal to the bit amount of the truth table ie. $k = \text{log}_2(N)$.\\

The metrics of interest are the average CNOT count and the average amount of arbitrary unitary rotations contained in the individual circuits. The latter will be denoted as $U$ count from now on and  is obtained by decomposing all occurring single-qubit gate instructions into their elementary rotations, that is:
\begin{align}
U(\theta,\phi,\lambda) = 
\begin{pmatrix}
\cos{\frac{\theta}{2}} &  -e^{i\lambda} \sin{\frac{\theta}{2}} \\
e^{i\phi} \sin{\frac{\theta}{2}} & e^{i(\phi+\lambda)} \cos{\frac{\theta}{2}}
\end{pmatrix} .
\end{align}
The duration and amount of the experiments (30 per datasize) are selected as to provide a reasonable understanding regarding the efficiency of
the methods under test. Hence, we have kept the sample fairly small and have not followed the path of statistical testing with the goal
to show any statistical significance.
The measured gate counts are rounded correspondingly to obtain reasonable results. Furthermore, we additionally show the 
savings in terms of CNOT counts by using the discussed solutions to the Hamming TSP instead of Gray-code in the phase tolerant 
synthesis. Those savings are more apparent in the case of smaller data sizes and smooth out when considering larger data 
sets. This does not have an impact on the $U$ counts and the decrease in quantum resources comes of course at the cost of an 
increase in classical resources. When comparing our results to \textit{tweedledum}'s Gray synthesis implementation, 
we consider the CNOT counts obtained by the corresponding implementation of phase tolerant synthesis with the results obtained through the Hamming TSP based approach.


As can be seen from Table \ref{Comparison_PTS_GRAY}, the comparison of both synthesis methods shows that the phase tolerant 
synthesis performs twice as good in terms of CNOT count when compared to the Gray synthesis implemented 
in \textit{tweedledum}. This is an impressive result that is confirmed after comparing the $U$ counts for both 
methods. For a database with $2^n$ entries, our phase tolerant synthesis as well as the standard Gray synthesis scale linear with the database size, which is an important property with regard to real world use cases. 
In the context of such real world use cases, it is not enough to look only at the general $U$ counts. Another interesting metric that can be extracted from the elementary single qubit rotations is the number of $\text{T}_m$ gates together with the T-order mentioned above. 
For this, we search for the $U_1$ rotations from the circuit data and extract the belonging angular parameters. These 
parameters are then represented in the most efficient sequence of $\text{T}_m$ gates. Thus, we can count all $\text{T}_m$ gates and compute 
the T-order corresponding to the largest occurring $m$ for each circuit. The corresponding data for both synthesis methods 
is shown in Table \ref{Comparison_PTS_GRAY}.


In the previous section, we addressed the implementation of scalable logic synthesis with the assumption that the 
number of qubits will likely increase more than gate precision and circuit depth for upcoming quantum hardware. The data 
above shows a linear increase in the T-order with increasing data size together with rapidly growing $\text{T}_m$ counts. 
With this in mind, we test a last synthesis method using the common sub-expression elimination technique (CSE) as described 
before in a similar test scenario. The obtained results are depicted in table \ref{Comparison_AVG_METRICS_CSE_SYNTH}. As the qubit resources now depend on the modified truth tables, additionally an average qubit count is given as well.

\begin{table*}[t]
    \centering
    \begin{tabular}{llllllllll}
        \toprule
        \multirow{2}{*}[-1em]{Data size} & \multicolumn{8}{c}{Average metrics} \\
        \addlinespace[10pt]
        \cmidrule(lr){2-9}\\
        {} & CNOT count & Comparison &  $U$ count & Comparison & Qubits & Comparison & $\text{T}_m$ count (T-order)& Comparison\\
        \midrule
        4 & 11  & 83.3\% & 8 & -11\%  & 9 & 125.0\%  &4 (2) & -33.3\%\\
        8 & 31 & 63.2\%  & 24 & 4\% & 14 & 133.3\%  & 16 (2) & -30.4\%\\
        16 & 69 & 16.9\%  & 63 & 5\% & 21 & 162.5\%  &41 (2) & -34.9\%\\
        32 & 151 & 2.0\%  & 150 & 1\% & 34 & 240.0\%  &100 (2) & -47.1\%\\
        64 & 316 & -13.4\% & 344 & -5\% & 56 & 366.7\%  &221 (2)& -59.6\%\\
        \bottomrule
    \end{tabular}
    \caption{Average metrics for the CSE-synthesis. The comparisons are described by procentual increases/decreases with respect to the gate counts 	obtained from the benchmark of our phase tolerant synthesis using HTSP solutions.}
		\label{Comparison_AVG_METRICS_CSE_SYNTH}
\end{table*}

We can only see a noticable decrease in CNOT counts and $U$ counts at a datasize of 64, whereas the qubit resources are 
several times larger. Since the qubit resources now depend on the modified truth tables and hence underlie some degree of 
variation, we have given the average results. For smaller data sizes, the $U$ counts are roughly on the same level as in the non CSE-supported synthesis methods 
and the CNOT counts have slightly increased in most of the analyzed cases. The main motivation for implementing the CSE 
method was to decrease the T-order and $\text{T}_m$ counts, which has proven to be highly successful as the T-order was kept 
constantly at 2 and the $\text{T}_m$ counts were reduced by up to 33\% . Indeed, we traded a decrease in gate complexity through 
fewer and considerably less fine phase gates at the cost of an increase in the qubit count and classical 
resources\footnote{This means that we are using more qubits and that the computations of the quantum logic synthesis takes longer on the classical computer.} for 
potential benefits in real-word applications.

\subsection{Comparison with Qiskits TruthTableOracle}

\begin{table*}[t]
    \centering
    \begin{tabular}{lllllllllllll}
        \toprule
        \multirow{2}{*}[-1em]{Data size} & \multicolumn{4}{c}{Average CNOT count} & \multicolumn{1}{c}{} & \multicolumn{4}{c}{Average $U$ count} &  \multicolumn{1}{c}{} & \multicolumn{2}{c}{Qubits}\\
        \addlinespace[10pt]
        \cmidrule(lr){2-5} \cmidrule(lr){7-10} \cmidrule(lr){12-13} \\
        {} & "basic" & Comparison & "noancilla" & Comparison & {} & "basic" & Comparison & "noancilla" & Comparison & {} & "basic" & "noancilla" \\
        \midrule
        4 & 9  &  50.0\% & 12  &  100.0\% & & 16 & 77.8\% & 24 & 166.7\% & & 4 & 4 \\        
        8 & 63 & 231.6\% & 84 & 342.1\% & & 113  &  391.3\% & 124 & 439.1\% & & 7 & 6 \\
        16 & 292 & 394.9\% & 420 & 611.9\% & & 597 & 895.0\% & 786 & 1210.0\% & & 10 & 8 \\
        32 & 960 & 548.6\% & 2722 & 1739.2\% & & 1993 & 1237.6\% & 3551 & 2283.2\% & & 13 & 10 \\
        64 & 2590 & 609.6\% & 13803 & 3681.6\% & & 5354 & 1374.9\%  & 15144 & 4071.9\% & & 16 & 12 \\
       \bottomrule
   \end{tabular}
    \caption{Average metrics for the \textit{TruthTableOracle} with the "basic" and "noancilla" option. The comparisons are described by procentual increases/decreases with respect to the gate counts 	obtained from the benchmark of our phase tolerant synthesis using HTSP solutions.}
		\label{AVG_TruthTableOracle}
\end{table*}

As a comparison of the complexity of arbitrary synthesized Grover oracles, we next check our \textit{phase tolerant Gray synthesis} method against the \textit{TruthTableOracle} class implemented in Qiskit Aqua. \footnote{At the time of preparing the current paper, this class is implemented in Qiskit Aqua 0.9.1 but due to a migration mainly to Qiskit Terra this is likely to change in the future.}. A brief description of this method can be seen in the official documentation and 
the source code is publicly available in the corresponding GitHub repository \cite{qiskit_Aqua}. 

The Qiskit-synthesized circuits can be further optimized by first minimizing the input truth table via the 
Quine-McCluskey\footnote{The Quine-McCluskey procedure is a method for minimizing logical/boolean formulas based on a standard 
representation and the identification and elimination of redundant terms.} 
algorithm \cite{Quine_McCluskey_Method}, in order to find all the essential 
prime implicants and then by finding the exact cover via employing the DLX\footnote{The DLX algorithm is an approach to 
solving the \textbf{set cover problem} through the methodology of dancing links \cite{Knuth2000} and was initially developed 
by Donald Knuth.} algorithm \cite{Knuth2000}. 
This is exponentially heavy on classical resources and it turns out that even with 
those optimizations, there is a huge gap in terms of CNOT counts and $U$ counts. The qubit count can be additionally 
addressed through a given selection of options concerning the implementation of multi-controlled Toffoli gates, which have a 
noticeable impact on the CNOT counts and $U$ counts as well. Hence, even when applying those optimizations for 
the Qiskit methods, we still get the same benchmarking results for the comparison between the different approaches.

We used the same test scenario as for the comparison of the synthesis methods above, however due to memory limitations\footnote{We ran 
the benchmark with a system memory of 32GB.} during the synthesis and optimization process, we have limited the range in data 
size to a maximum of 64 data entries. The mode for constructing multi-controlled Toffoli gates was first set to the "basic" 
setting. The results for the CNOT counts and $U$ counts can be seen in Table \ref{AVG_TruthTableOracle}.


The difference in circuit complexity between the \textit{TruthTableOracle} and our method is in significant favour
of our method and even more so, when considering the necessary number of qubits for the oracle circuits. 
The "basic" option for multi-controlled 
Toffoli construction delivers the best scaling behavior in terms of gate counts while using $3n - 2$ qubits for the tested 
data with $2^n$ entries, as opposed to our method using only $2n$ qubits. The option leading to the lowest number of 
qubits - which is also given by $2n$ - is the "noancilla" one. This method leads to a much worse scaling behavior as can be 
seen in Table \ref{AVG_TruthTableOracle} below. 


We also want to mention, that there is another implementation in Qiskit's \textit{PhaseOracle} class, which takes a boolean expression or 
a SAT problem in DIMACS CNF format \cite{DIMACS_CNF} and passes this to the PPRM
synthesis method implemented in \textit{tweedledum}. Thus, we don't give a further comparison between our method and the 
\textit{PhaseOracle} class as we have tested the different synthesis methods implemented in \textit{tweedledum} and identified
the Gray synthesis as the most suited method and competitor in the  context of array oracle generation for Grover's 
algorithm.

\label{sec:summary}
\section{Summary}
\label{summary}

The current paper provides a first successful attempt for enabling the convenient utilization of
Grover's search algorithm capabilities over traditional function/procedure APIs (e.g.
\textit{int grover(int [] list\_to\_search\_in, int value\_to\_search\_for)}). The motivation
for this research is based on observations in our previous work \cite{GheorghePop2021}, in which
the issues of constructing oracles for Grover database searches was discussed from a user perspective.
Indeed, the only way to make quantum computing commercially viable is to provide accessible interfaces for programmers and end users to integrate quantum algorithms in their
services and applications.

With regard to Grover's algorithm, such APIs require for the automatic generation
of the black box quantum oracles, which contain the database and the element to
search for in this database. In this context, our current paper provides a methodology 
for automatically generating such quantum oracles for arbitrary databases. The generation
consists of two main parts: 1) Mapping the database entries to a circuit $U_D$ generated by logic synthesis and 2)
tagging the query hash to create the query oracle. The first step is realized through
the utilization of beyond state-of-the-art synthesis functions, while the second step can be realized either with the traditional multi-controlled Z gates or with our newly introduced similarity tags. In this regard, one of the main contributions
of this paper is given by the phase tolerant enhancement of synthesis procedures, allowing for resource cuts up to 50\% within the context of Grover quantum oracle generation. Furthermore, we present a new synthesis method respecting the requirements of scaling the synthesis procedure for real world physical backends.\\
To summarize: this paper outlines a clear procedure for making the potentials of the powerful quantum algorithm by Grover available to programmers and end users for integration in
everyday ICT-systems (e.g. online shops, telecommunication management systems, database search engines,
web analytic systems ...). The methodology proposed in this paper generates the belonging quantum
oracles automatically, thereby utilizing and leading to innovative methods for quantum logic synthesis.
The computational complexity of the methodology is in general higher than the one of classical search. However, our future research works aims at optimizing this complexity through different 
heuristics, machine learning techniques and optimizations on the proposed approach. By continuously achieving such
gradual improvements, one can see a clear path to a full-scale introduction and application of quantum algorithms based on oracles in current development processes and system architectures.

\bibliographystyle{IEEEtran}
\bibliography{Introduction_sources}

\vfill


\end{document}